\documentclass[prd,preprint,tightenlines,floatfix,showpacs,%
preprintnumbers,nofootinbib,eqsecnum]{revtex4}
\usepackage[dvips,final]{graphicx}
\usepackage{epsfig}
\usepackage{bm}
\begin{document}

\preprint{\hbox{RUB-TPII-18/02}}
\date{\today}
\title{Power corrections to the space-like transition form factor
       ${\mathbf{F_{\eta^{\prime}g^{*}g^{*}}(Q^2,\omega )}}$ \\}
\author{\textbf{S.~S.~Agaev}}
\email{agaev_shahin@yahoo.com}
\thanks{Permanent address: High Energy Physics Lab.,
        Baku State University,
        Z. Khalilov St. 23, 370148 Baku, Azerbaijan}
\affiliation{The Abdus Salam International Centre
             for Theoretical Physics,
             I-34014 Trieste, Italy}
\author{\textbf{N.~G.~Stefanis}}
\email{stefanis@tp2.ruhr-uni-bochum.de}
\affiliation{Institut f\"ur Theoretische Physik II,
             Ruhr-Universit\"at Bochum,
             D-44780 Bochum, Germany}

\begin{abstract}
Employing the standard hard-scattering approach (HSA) in conjunction
with the running-coupling (RC) method, the latter joined with the
infrared-renormalon calculus, we compute power-suppressed corrections
$\sim 1/Q^{2n},\,\,\,n=1,\,2,...$ to the massless
$\eta^{\prime}$ meson - virtual gluon transition form factor (FF)
$Q^2F_{\eta ^{\prime }g^{*}g^{*}}(Q^2,\omega)$.
Contributions to the form factor from the quark and gluon components
of the $\eta^{\prime}$ meson are taken into account.
Analytic expressions for the FF's
$F_{\eta^{\prime}gg^{*}}(Q^2,\omega=\pm 1)$
and $F_{\eta^{\prime}g^{*}g^{*}}(Q^2,\omega=0)$
are also presented, as well as Borel transforms
$B[Q^{2}F_{\eta^{\prime}g^{*}g^{*}}](u)$
and resummed expressions.
It is shown that except for $\omega=\pm 1,\: 0$, the Borel transform
contains an infinite number of infrared renormalon poles.
It is demonstrated that in the explored range of the total gluon
virtuality $1 \, \rm {GeV}^{2} \leq Q^2 \leq 25 \,\rm {GeV}^{2}$,
power corrections found with the RC method considerably enhance the
FF $F_{\eta^{\prime}g^{*}g^{*}}(Q^{2}, \omega)$ relative to results
obtained only in the context of the standard HSA with a ``frozen''
coupling.
\end{abstract}

\vspace{0.5 cm}
\hspace{3 cm}

\maketitle
\newpage

\section{Introduction}
\setcounter{equation}0
During the last few years the interest into theoretical investigations
of the quark-gluon structure of light mesons, especially the pion,
$\eta$ and $\eta^{\prime}$ mesons, has risen due to the high-precision
CLEO results on the electromagnetic $M\gamma$ transition form factors
(FF's) $F_{M\gamma}(Q^2)$ \cite{cleo1} (with $M$ denoting one of these
mesons), as well as because of the observed very large branching ratios
for the exclusive $B\rightarrow K\eta^{\prime}$ and semi-inclusive
$B\rightarrow \eta^{\prime}X_s$ decays \cite{cleo2}.

The data on the $\eta^{\prime}\gamma$ transition FF were mainly used
for extracting information concerning the $\eta^{\prime}$ quark
component of the meson distribution amplitude (DA).
Schemes and methods applied to this purpose range from light-cone
perturbation-theory calculations -- with the quark transverse-momentum
$k_{\perp}$ dependence kept in the hard-scattering amplitude
$T_H(x,k_{\perp},Q^2)$ of the underlying hard subprocess \cite{cao} --
to the modified hard-scattering approach (mHSA) with resummed
transverse-momentum effects (giving rise to Sudakov suppression
factors) and such due to the intrinsic transverse momentum of the
meson wave functions \cite{jakob,feld1}, to the running coupling (RC)
method, employed for the estimation of power-suppressed corrections to
the $\eta^{\prime}\gamma$ transition FF \cite{ag1}.
In these investigations, two different parameterizations were used, one
employing the conventional $\eta -\eta^{\prime}$ mixing scheme with one
mixing angle $\theta_p$ \cite{cao,jakob,ag1} for both physical states
and decay constants and a second one, which considers two mixing angles
$\theta_1$ and $\theta_8$ to parameterize the weak decay constants
$f_P^i$ $(P=\eta,\,\eta^{\prime};\,\,i=1,8)$ of the $\eta$
and $\eta^{\prime}$ mesons \cite{feld1}.
An important conclusion drawn from these investigations, irrespective
of the underlying method, is that the $\eta^{\prime}$-meson DA must be
close to its asymptotic form and that the admixture of the first
non-asymptotic term should be within the range
$b_2(\mu_0^2)\simeq 0.05 \div 0.15$, $b_2$ being the first Gegenbauer
coefficient.
The CLEO data on the $\eta^{\prime}\gamma$ transition and the
two-angles mixing scheme were also used to estimate the allowed range
of the intrinsic charm content of the $\eta^{\prime}$ meson decay
constant $f_{\eta^{\prime}}^c$ \cite{feld1}.
It turned out that the value
$-65\,\,\,\rm{MeV}\leq $$f_{\eta^{\prime}}^c\leq 15\,\,\rm{MeV}$
does not contradict the CLEO data.

But apart from the ordinary light quark and charm
$\left|\eta^{\prime}_c\right\rangle$
components, the $\eta^{\prime}$ meson may also contain a two-gluon
valence Fock state $\left|gg\right\rangle$.
Moreover, absent at the normalization point $\mu_0$, a gluon component
of the $\eta^{\prime}$ meson will appear in the region $Q^2>\mu_0^2$
owing to quark-gluon mixing and renormalization-group evolution of the
$\eta^{\prime}$ meson DA \cite{chase1,ter81,ohrndorf,shif,ba}.
This can directly contribute to the $\eta^{\prime}\gamma$ transition at
the next-to-leading order due to quark box diagrams and also affect the
leading-order result through evolution of the quark component of the
$\eta^{\prime}$-meson DA.
Hence, an effect of the $\eta^{\prime}$-meson gluon component on the
$\eta^{\prime}\gamma$ transition is only mild and was therefore
neglected in most theoretical investigations
\cite{cao,jakob,feld1,ag1}.

But the contribution of the gluon content of the $\eta^{\prime}$ meson
to the two-body non-leptonic exclusive and semi-inclusive decay ratios
of the $B$ meson may be sizeable.
Indeed, such a mechanism to account for the observed large branching
ratio \cite{cleo2}
\begin{equation}
\label{eq:1.1}Br(B\rightarrow \eta ^{\prime }+X_s)=(6.2\pm 1.6\pm 1.3)
\times
10^{-4},
\end{equation}
was proposed in Ref.\ \cite{at}.
In this work it was suggested that the dominant fraction of the
$B\rightarrow \eta^{\prime}+X_s$ decay rate appears as the result of
the transition $g^{*}\rightarrow g\eta$ of a virtual gluon $g^{*}$ from
the standard model penguin diagram $b\rightarrow sg^{*}$.
For the computation of the contribution of this mechanism to the
$Br(B \to \eta^{\prime}+X_s)$ in Ref.\ \cite{at}, the
$g^{*}g\eta^{\prime}$ vertex function was approximated by the constant
$H(q^2,0,m_{\eta^{\prime}}^2)\simeq H(0,0,m_{\eta^{\prime}}^2)
\simeq 1.8~{\rm GeV}^{-1}$, the latter being  extracted from the
analysis of the $J/\psi \rightarrow \eta^{\prime}\gamma$ decay.
Further investigations, however, demonstrated that effects of the QCD
running coupling $\alpha_{\rm s}(q^2)$ \cite{hou}, as well as a
momentum dependence of the form factor $H(q^2,0,m_{\eta^{\prime}}^2)$,
properly taken into account, considerably reduce the contribution to
Eq.\ (\ref{eq:1.1}) of the mechanism under consideration \cite{kag}.
In order to cover the gap between theoretical predictions and
experimental data in Ref.\ \cite{ah}, a gluon fusion (spectator)
mechanism was proposed.
In accordance with the latter, the $\eta^{\prime}$ meson is produced by
the fusion of a gluon from the QCD penguin diagram
$b\rightarrow sg^{*}$ with another one emitted by the light quark
inside the $B$ meson.
In this mechanism, the form factor \footnote{In this work we use the
notions ``vertex function'' and ``form factor'' on the same footing.}
$F_{\eta^{\prime}g^{*}g^{*}}(q{_1}^2,q_{2}^2,m_{\eta^{\prime}}^2)$
appears owing to the $g^{*}g^{*}\rightarrow \eta^{\prime}$ transition.

The same ideas form a basis for the computation of the branching ratios
of various two-body non-leptonic exclusive decay modes of the $B$ meson
\cite{ali1}.
To account for the data on the exclusive decay
$B\rightarrow K\eta^{\prime}$ in Ref.\ \cite{hal}, a mechanism was
proposed, based on the assumption of a strong intrinsic charm component
of the $\eta^{\prime}$ meson.
But a more detailed analysis \cite{franz} proved that the charm content
of the $\eta^{\prime}$ is too small
($f^{c}_{\eta^{\prime}}\simeq -2~\rm{MeV}$) to explain the observed
branching ratio $Br(B \rightarrow K \eta^{\prime})$.
The CLEO data on the $B$ meson non-leptonic decays that stimulated
interesting theoretical works \cite{dihle} remain actual until today
\cite{eeg}.

The exclusive processes $B\to K^{(*)}\eta^{(\prime)}$ were also
analyzed within the QCD factorization approach \cite{neub1} and various
contributions to the corresponding branching ratios were estimated
\cite{neub2}.
In accordance with Ref.\ \cite{neub2}, the spectator mechanism and
such connected to the gluon content of the $\eta$ and $\eta^{\prime}$
mesons are not a key factor in explaining the pattern of the observed
experimental data.
Instead, a more important role plays here the interference
(constructive or destructive) of penguin amplitudes and large radiative
corrections to them, which bring the predicted branching ratios into
reasonable agreement with the data.

The $\eta^{\prime}$-meson virtual-(on-shell-)gluon transition form
factor $F_{\eta^{\prime}g^{*}g^{*}}(q_1^2,q_2^2,m_{\eta^{\prime}}^2)$
is the central ingredient of analyses performed within perturbative
QCD (pQCD) and hence requires further thorough investigations.
Such theoretical investigations are especially important in view of
contradictory predictions made for this FF in the literature
\cite{muta,yang,ali2,pas}.
For instance, in Ref.\ \cite{muta} the contribution of the gluon
content of the $\eta^{\prime}$ meson to the form factor
$F_{\eta^{\prime}gg^{*}}(q_1^2,0,m_{\eta^{\prime}}^2)$
was estimated and found to be too small while
$F_{\eta^{\prime}gg^{*}}(q_1^2,0,m_{\eta^{\prime}}^2)$
itself is close to the form
\begin{equation}
\label{eq:1.2}F_{\eta^{\prime}gg^{*}}(q_1^2,0,m_{\eta^{\prime}}^2)
\simeq
\frac{H(0,0,m_{\eta^{\prime}}^2)}{q_1^2/m_{\eta^{\prime}}^2-1}
=\frac{1.8\,\,}{q_1^2/m_{\eta^{\prime}}^2-1}\,\,\,\rm{GeV}^{-1},
\end{equation}
used in some phenomenological applications \cite{kag,ah}.
On the other hand, in computations of the $\eta^{\prime}$-meson
virtual-gluon vertex function
$F_{\eta^{\prime}g^{*}g^{*}}(q_1^2,q_2^2,m_{\eta^{\prime}}^2)$,
the gluon content of the $\eta^{\prime}$ meson was neglected and the
asymptotic form of the DA for the quark component was employed
\cite{yang}.
The same vertex function
$F_{\eta^{\prime}g^{*}g^{*}}(q_1^2,q_2^2,m_{\eta^{\prime}}^2)$
was considered in \cite{ali2}, where both the standard HSA \cite{br1}
as well as the mHSA were used, and the space-, and time-like vertex
functions were analyzed.
Some errors made in Ref.\ \cite{muta} in the description of the
hard-scattering amplitude of the subprocess $g + g \to g^{*} + g^{*}$
and in the definition of the evolution of the $\eta^{\prime}$ meson DA
were corrected in \cite{ali2}.
Unfortunately, also this latter work contains an incorrect expression
for the gluon component
$F^{g}_{\eta^{\prime}g^{*}g^{*}}(Q^2, \omega, \eta)$
of the form factor, since it is antisymmetric under the exchange of the
asymmetry parameter $\omega \leftrightarrow -\omega$ in
clear conflict with Bose symmetry.
More recently, in Ref.\ \cite{pas}, the space-like form factor
$F_{\eta^{\prime}g^{*}g^{*}}(Q^2,\omega)$ was computed within the
standard HSA.
These authors performed a detailed analysis of the normalization of
the gluon component of both the $\eta^{\prime}$-meson DA and that of the
gluon projector onto a pseudoscalar meson state.

In the present work, we investigate the massless $\eta^{\prime}$-meson
virtual-gluon space-like transition form factor
$F_{\eta^{\prime}g^{*}g^{*}}(Q^2,\omega)$
using the framework of the standard HSA \cite{br1}, as well as by applying
the RC method together with the infrared (IR) renormalon calculus
\cite{ben}.
Our results obtained within the standard HSA are in agreement with those
of Ref.\ \cite{pas}.
But our central task here is the calculation and evolution of
power-suppressed corrections
$\sim 1/Q^{2n},\,\,n=1,2...$ to the transition FF
$Q^2F_{\eta^{\prime}g^{*}g^{*}}(Q^2,\omega )$.
Because in the production of the $\eta^{\prime}$ meson from the $B$
decay the momentum squared of the virtual gluon can vary from
$1\;\,\rm{GeV}^2$ to $25\,\,\rm{GeV}^2$, the power corrections in this
domain of $Q^2$ are expected to play an important role.
Note, however, that we elaborate only on a theoretical framework to
compute power corrections for the space-like transition FF.
Nevertheless, our technique can be generalized to encompass the
time-like transition form factor, relevant for B meson decays,
as well.
Work in this direction will be reported elsewhere.

The RC method enables us to estimate power corrections coming from the
end-point $x,\,y\rightarrow 0,1$ regions (for definiteness we consider
two mesons in an exclusive process) in the integrals determining the
amplitude for an exclusive process.
It is known \cite{br1} that in order to calculate an amplitude of some
hadron exclusive process, one has to perform integrations over
longitudinal momentum fractions $x,\,\,y$ of the involved partons.
If one chooses the renormalization scale $\mu_R^2$ in the
hard-scattering amplitude $T_H$ of the corresponding partonic
subprocess in such a way as to minimize higher-order corrections and
allows the QCD coupling constant $\alpha_{\rm s}(\mu_R^2)$ to run,
then one encounters divergences arising from the end-point
$x,\,y\rightarrow 0,1$ regions.
The reason is that the scale $\mu_R^2$, as a rule, is equal to the
momentum squared of the hard virtual partons, the latter carrying the
strong interactions in the subprocess' Feynman diagrams \cite{br2}
and, in general, depends on $x$ and $y$.
Within the RC method this problem is resolved by applying the IR
renormalon calculus.
It turns out that this treatment allows us to evaluate power-behaved
corrections to the physical quantity under investigation.
This method was recently used for the computation of power corrections
to the electromagnetic form factors $F_{\pi ,K}(Q^2)$ of the pion and
the kaon, \cite{ag2,ag3}, and the electromagnetic transition FF's
$F_{M\gamma}(Q^2)$ \cite{ag1,kivel} of the light pseudoscalar mesons.
Power corrections to hadronic processes can also be calculated
utilizing the Landau-pole free expression for the QCD coupling constant
\cite{shir}.
This analytic approach was used to compute in a unifying way power
corrections to the electromagnetic pion form factor and to the
inclusive cross section of the Drell-Yan process \cite{stef,stef1}.

The paper is organized as follows.
In Sect.\ II we briefly review $\eta-\eta^{\prime}$ mixing schemes in
the flavor $SU_f(3)$ octet-singlet and the quark-flavor bases, and
discuss the evolution of the quark and gluon components of the
$\eta^{\prime}$ meson DA with the factorization scale.
The important question of the generalization of the hard-scattering
amplitudes of the $\eta^{\prime}g^{*}$ transition to the RC method case
is also considered.
Section III is devoted to a rather detailed presentation of the RC
method.
In Sect.\ IV  we compute the quark and gluon components of the
transition
FF $F_{\eta^{\prime}g^{*}g^{*}}(Q^2,\omega)$.
Section V contains our numerical results.
Finally, in Sect.\ VI we make our concluding remarks.
Appendix contains an additional information on the $\eta^{\prime}$-
meson DA.
\section{Quark and gluon contributions to the
         $\mathbf{\eta^{\prime}-g^{*}}$
         transition form factor}
\setcounter{equation}0
In this section we consider the quark-gluon content of the
$\eta^{\prime}$ meson, as well as the $\eta^{\prime}$ DA and give some
general expressions for determining the $\eta^{\prime}g^{*}$
transition form factor within both the standard HSA and the RC method.

\subsection{Structure of the $\mathbf{\eta^{\prime}}$ meson}

The parton Fock state decomposition of the pseudoscalar
$P=\eta,\,\eta^{\prime}$
mesons, can be generically written in the following form
\begin{equation}
\label{eq:2.1}
\left\vert P\right\rangle
= \left\vert P_a \right\rangle
+\left\vert P_b \right\rangle
+\left\vert P_c \right\rangle
+\left\vert P_g \right\rangle,
\end{equation}
where $\left\vert P_a \right\rangle$ and $\left\vert P_b \right\rangle$
denote the
$P$-meson light-quarks $u,\,d,\,s$ and
$\left\vert P_c \right\rangle$, $\left\vert P_g \right\rangle$ its charm
and gluon components, respectively.
Fock states with additional gluons and $q\overline q$ quark-antiquark
pairs have been omitted for simplicity.

The $P$-meson light-quark content
$\left\vert P_a \right\rangle,\, \left\vert P_b \right\rangle$
can be described either in the flavor $SU_{f}(3)$ octet-singlet or in
the quark-flavor basis.
In the first scheme, the states
$\left\vert P_a \right\rangle,\, \left\vert P_b \right\rangle$
are expressed as superpositions of the $SU_{f}(3)$ singlet $\eta_1$ and
octet $\eta_8$ states
$$
\left\vert\eta_1\right\rangle
=
\frac{\Psi_1}{\sqrt{3}}
\left\vert u\overline{u}+d\overline{d}+s\overline{s}\right\rangle ,
$$
\begin{equation}
\label{eq:2.2}
\left\vert\eta_8\right\rangle = \frac{\Psi_8}{\sqrt{6}}\left\vert
u\overline{u}+d
\overline{d}-2s\overline{s}\right\rangle ,
\end{equation}
whereas in the quark-flavor basis, the $SU_{f}(3)$ strange $\eta_s$ and
non-strange $\eta_q$ states are used, i.e.,
\begin{equation}
\label{eq:2.3}
\left\vert\eta_q\right\rangle = \frac{\Psi_q}{\sqrt{2}}\left\vert
u\overline{u}+d
\overline{d}\right\rangle ,\,\,\,\,\,\left\vert\eta_s\right\rangle
=\Psi{_s}
\left\vert
\overline{s}\right\rangle .
\end{equation}
In Eqs.\ (\ref{eq:2.2}) and (\ref{eq:2.3}) $\Psi_i$ denote wave
functions of the corresponding parton states.

As mentioned above, the charm component of the $\eta^{\prime}$ meson
was estimated \cite{franz,pet} to be too small to considerably affect
the $B$ meson exclusive decays.
Therefore, in the present investigation we neglect the charm content
of the $\eta^{\prime}$ meson.
The maximal admixture of the two-gluon state in the $\eta^{\prime}$
meson was estimated to be around $26\%$ of its content \cite{kou}.
New results from the KLOE Collaboration \cite{aloisio} are compatible
with the two-gluon contribution in the $\eta^{\prime}$ meson being
below the level of $15\%$.

The pure light-quark sector of the $\eta - \eta^{\prime}$ system
without charm and gluon admixtures can be treated by means of the basic
states (\ref{eq:2.2}) or (\ref{eq:2.3}).
In the $SU_{f}(3)$ octet-singlet basis, the $\eta$ and $\eta^{\prime}$
mesons are expressed as superpositions of the $\eta_8,\,\eta_1$ states
$$
\left| \eta \right\rangle =\cos \theta_p \left| \eta _8\right\rangle
-\sin\theta_p \left| \eta _1\right\rangle ,
$$
\begin{equation}
\label{eq:2.5}
\left| \eta ^{\prime }\right\rangle =\sin \theta_p \left| \eta
_8\right\rangle
+\cos \theta_p \left| \eta _1\right\rangle .
\end{equation}
Here, $\theta_p $ is the mixing angle of physical states in the
octet-singlet scheme.
In the quark-flavor basis we get the same expressions but with
$\theta_p$ replaced by the mixing angle $\phi_p$ of the physical
states in the new basis
$\eta_q,\,\eta_s$, viz.,
$$
\left| \eta \right\rangle =\cos \phi_p \left| \eta _q\right\rangle
-\sin\phi_p
\left| \eta _s\right\rangle ,
$$
\begin{equation}
\label{eq:2.6}
\left| \eta ^{\prime }\right\rangle =\sin \phi_p \left| \eta
_q\right\rangle
+\cos \phi_p \left| \eta _s\right\rangle .
\end{equation}

The $\eta _1$ and $\eta _8$ states and the mixing angle $\theta_p $ in
the octet-singlet scheme can be expressed in terms of the
$\eta _q,\,\,\eta_s$ states and the mixing angle $\phi_p $ in the
quark-flavor basis and vice versa.
At the level of physical-state mixing there is no difference
between the two bases (\ref{eq:2.2}) and (\ref{eq:2.3}).
This only appears when one parameterizes the decay constants
$f_P^i$ of the $P=\eta $ and $\eta ^{\prime }$ mesons in terms of
$$
\left \langle 0 \vert J_{\mu 5}^i \vert P \right \rangle =if_P^ip_\mu ,
$$
where $J_{\mu 5}^i$ is the axial-vector currents with
$i=q,\,s$ or $i=1,\,8$.
In the quark-flavor basis the decay constants $f_P^{q(s)}$ follow with
great accuracy the pattern of the state mixing \cite{feld2,feld3}
$$
f_\eta ^q=f_q\cos \phi_p ,\,\,\,f_\eta ^s=-f_s\sin \phi_p ,
$$
\begin{equation}
\label{eq:2.7}
f_{\eta ^{\prime }}^q=f_q\sin \phi_p ,\,\,\,f_{\eta ^{\prime }}^s
=f_s\cos\phi_p.
\end{equation}
The situation with the parameterization of the decay constants
in the octet-singlet basis is different. In this case, in order to
take into account the flavor $SU_f(3)$ symmetry breaking effects, a
two mixing-angles scheme for the decay constants $f_P^{1(8)}$ was
introduced \cite{let}
$$
f_\eta ^8=f_8\cos \theta _8,\,\,\,f_\eta ^1=-f_1\sin \theta _1,
$$
\begin{equation}
\label{eq:2.8}
f_{\eta ^{\prime }}^8=f_8\sin \theta _8,\,\,\,f_{\eta ^{\prime
}}^1=f_1\cos
\theta _1.
\end{equation}
The mixing angles $\theta _1$ and $\theta _8$ differ from each other
and also from the state mixing angle $\theta_p $.
Nevertheless, in some phenomenological applications the conventional
octet-singlet mixing scheme, that is the scheme with one mixing
angle and which assumes the equality $\theta _1=\theta_8=\theta_p$,
is used.
In the investigation of physical processes both the octet-singlet and
the quark-flavor bases may be used.
A detailed analysis of the $\eta - \eta^{\prime}$ mixing problems and
further references can be found in Ref.\ \cite{feld4}.

In our present work we choose the quark-flavor basis (\ref{eq:2.3}) and
the mixing scheme (\ref{eq:2.6}).
In this scheme, the decay constants $f_q$ and $f_s$, and the mixing
angle $\phi_p$ have the following values \cite{feld3,feld4}
$$
f_q=(1.07 \pm 0.02)f_{\pi},\,\,f_s=(1.34 \pm 0.06)f_{\pi},\,\,
\phi_p=39.3^{\circ} \pm 1.0^{\circ}
$$
with $f_{\pi}=0.131\,\rm{GeV}$ being the pion weak decay constant.
The KLOE result for
$\phi_p=(41.8^{\circ +1.9^{\circ}}_{-1.6^{\circ}})$ \cite{aloisio}
is slightly shifted towards larger values, but it still does not
contradict the average value
$\phi_p=39.3^{\circ} \pm 1.0^{\circ}$.
In our numerical computations we shall use the central values of the
constants shown above.

The singlet part of the $\eta^{\prime}$ meson DA\footnote {The octet
component of the $\eta^{\prime}$ meson DA is irrelevant for our present
investigation and will not be considered here.
In what follows we refer to the singlet part of the $\eta^{\prime}$
meson DA as being the $\eta^{\prime}$ meson DA. } $\phi (x,\mu ^2)$
was obtained in Refs.\ \cite{ohrndorf,shif,ba} by solving the evolution
equation and found to depend on both functions $\phi^q(x,\mu^2)$ and
$\phi^g(x,\mu^2)$.
These functions denote the quark and gluon components of the
$\eta^{\prime}$ meson DA, respectively, and satisfy the symmetry and
antisymmetry conditions,
\begin{equation}
\label{eq:2.10}\phi ^q(x,\mu ^2)=\phi ^q(1-x,\mu ^2),
\,\,\,\,\,\phi ^g(x,\mu
^2)=-\phi ^g(1-x,\mu ^2).
\end{equation}
This follows from the symmetry properties of the DA of the two-particle
bound state of a neutral pseudoscalar meson and is in fact enough to
obtain general expressions for the $\eta^{\prime}$-meson virtual-gluon
transition form factor.
The evolution of the functions $\phi^q(x,\mu^2)$ and $\phi^g(x,\mu^2)$
with the scale $\mu^2$, as well as their dependence on the constants
$f_q,\;f_s$ and $\phi_p$ will be considered in Subsect.\ II.C.

\subsection{The $\mathbf{\eta^{\prime}-g^{*}}$ transition form factor
            $\mathbf{F_{\eta^{\prime }g^{*}g^{*}}(Q^2,\omega)}$}
The massless $\eta^{\prime}$ meson - virtual gluon transition form
factor
$F_{\eta^{\prime }g^{*}g^{*}}(Q^2,\omega )$
\begin{equation}
\label{eq:2.11}F_{\eta ^{\prime }g^{*}g^{*}}(Q^2,\omega)=F_{\eta
^{\prime
}g^{*}g^{*}}^q(Q^2,\omega )+F_{\eta ^{\prime
}g^{*}g^{*}}^g(Q^2,\omega),
\end{equation}
can be defined in terms of the invariant amplitude
\begin{equation}
\label{eq:2.12}M=M^q+M^g,
\end{equation}
for the process
\begin{equation}
\label{eq:2.13}\eta ^{\prime }(P)\rightarrow g^{*}(q_1)+g^{*}(q_2),
\end{equation}
in the following way
\begin{equation}
\label{eq:2.14}
M^{q(g)}=-iF_{\eta ^{\prime }g^{*}g^{*}}^{q(g)}(Q^2,\omega
)\delta _{ab}
\epsilon ^{\mu \nu \rho \sigma }\epsilon _\mu ^{a*}\epsilon
_\nu ^{b*}q_{1\rho }q_{2\sigma }.
\end{equation}
In Eq.\ (\ref{eq:2.14})
$\epsilon _\mu ^a,\,\,\epsilon _\nu ^b$ and $q_1,\;q_2$ are,
respectively, the polarization vectors and four-momenta of the two
gluons.
Because we study only the space-like FF, $q_1^2$ and $q_2^2$ obey the
constraints
$Q_1^2=-q_1^2\geq 0,\,\,Q_2^2=-q_2^2>0$ and
$Q_1^2=-q_1^2>0,\,\,Q_2^2=-q_2^2
\geq 0$.
The form factor $F_{\eta ^{\prime }g^{*}g^{*}}(Q^2,\omega)$
depends on the total gluon virtuality $Q^2$ and the asymmetry parameter
$\omega$, defined by
\begin{equation}
\label{eq:2.15}Q^2=Q_1^2+Q_2^2,\,\,\,\omega =\frac{Q_1^2-Q_2^2}{Q^2}.
\end{equation}
The asymmetry parameter $\omega $ varies in the region $-1\leq \omega
\leq 1$.
The value $\omega =\pm 1$ corresponds to the $\eta ^{\prime }$-meson
on-shell-gluon transition and the value $\omega =0$ to the situation
when the gluons have equal virtualities $Q_1^2=Q_2^2$.

\begin{figure}
\epsfxsize=10 cm
\epsfysize=7 cm
\centerline{\epsffile{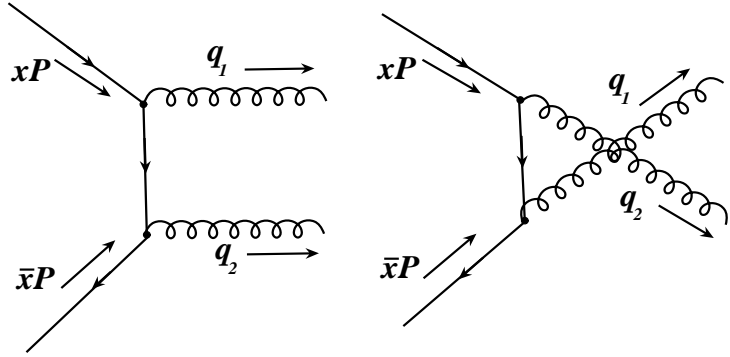}}
\caption{\label{fig:quark} Leading-order Feynman diagrams contributing
to the hard-scattering subprocess $q+\overline{q} \to g^{*}+g^{*}$.}
\end{figure}

\begin{figure}
\epsfxsize=10 cm
\epsfysize=7 cm
\centerline{\epsffile{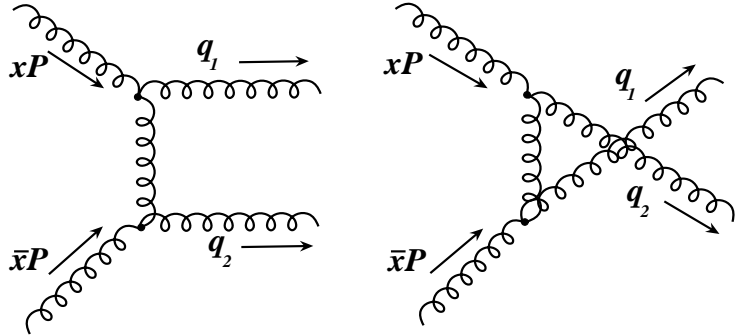}}
\caption{\label{fig:gluon}Feynman diagrams contributing at leading
order to the subprocess $g+g \to g^{*}+g^{*}$. }
\end{figure}

In accordance with the factorization theorems of pQCD at high momentum
transfer, the FF's $F_{\eta^{\prime}g^{*}g^{*}}^{q(g)}(Q^2,\omega )$
are given by the expressions
\begin{equation}
\label{eq:2.16}F_{\eta ^{\prime }g^{*}g^{*}}^q(Q^2,\omega)=\left[
T_H^q(x,Q^2,\omega ,\mu _F^2)+T_H^q(\overline{x},Q^2,\omega
, \mu _F^2)\right] \otimes \phi ^q(x,\mu _F^2)
\end{equation}
and
\begin{equation}
\label{eq:2.17}F_{\eta ^{\prime }g^{*}g^{*}}^g(Q^2,\omega)=\left[
T_H^g(x,Q^2,\omega, \mu _F^2)-T_H^g(\overline{x},Q^2,\omega,
\mu _F^2)\right] \otimes \phi ^g(x,\mu _F^2).
\end{equation}
Here, $\overline{x}\equiv 1-x$ and $\mu _F^2$ represents the
factorization scale.
In Eqs.\ (\ref{eq:2.16}) and (\ref{eq:2.17}) we have used the notation
\begin{equation}
\label{eq:2.18}T_H(x,Q^2,\omega, \mu _F^2)\otimes \phi (x,\mu
_F^2)=\int_0^1T_H(x,Q^2,\omega, \mu _F^2)\phi (x,\mu _F^2)dx.
\end{equation}
The sum
\begin{equation}
\label{eq:2.19}T_H^q(x,Q^2,\omega, \mu _F^2)+T_H^q(\overline{x}
,Q^2,\omega, \mu _F^2)
\end{equation}
and the difference
\begin{equation}
\label{eq:2.20}T_H^g(x,Q^2,\omega, \mu _F^2)-T_H^g(\overline{x}
,Q^2,\omega, \mu _F^2)
\end{equation}
represent the hard-scattering amplitudes of the subprocesses
$q+\overline{q}\rightarrow g^{*}+g^{*}$
and $g+g\rightarrow g^{*}+g^{*}$, respectively.
The Feynman diagrams contributing at the leading order to these
subprocesses' amplitudes are depicted in
Figs.\ \ref{fig:quark} and \ref{fig:gluon}.

In what follows, we omit the subscript $H$ in Eqs.\ (\ref{eq:2.19})
and (\ref{eq:2.20}) and introduce instead the notations
$$
T_1^{q(g)}(x,Q^2,\omega, \mu _F^2)=T_H^{q(g)}(x,Q^2,\omega,
\mu _F^2),
$$

$$
T_2^{q(g)}(x,Q^2,\omega , \mu _F^2)=T_H^{q(g)}(
\overline{x},Q^2,\omega , \mu _F^2).
$$

In leading order of pQCD, we get for the hard-scattering amplitudes
$T_1^{q(g)}$ and $T_2^{q(g)}$ of the massless $\eta^{\prime}$ meson -
virtual gluon transition FF the following expressions

$$
T_1^q(x,Q^2,\omega, \mu _R^2)=-\frac{4\pi }{3Q^2}\frac{\alpha_{\rm
s}(\mu
_R^2)}{(1+\omega )x+(1-\omega )\overline{x}},
$$

\begin{equation}
\label{eq:2.21}T_2^q(x,Q^2,\omega, \overline{\mu }_R^2)=-\frac{4\pi
}{3Q^2}
\frac{\alpha_{\rm s}(\overline{\mu }_R^2)}{(1+\omega
)\overline{x}+(1-\omega
)x}
\end{equation}
and

$$
T_1^g(x,Q^2,\omega, \mu _R^2)=\frac{\pi \alpha_{\rm s}(\mu
_R^2)}{Q^2 n_{f}}
\frac{(1+\omega )x+(1-\omega )\overline{x}}{\omega \left [
(1+\omega )\overline{x}+(1-\omega )x \right ]},
$$

\begin{equation}
\label{eq:2.22}T_2^g(x,Q^2,\omega, \overline{\mu }_R^2)=\frac{\pi
\alpha_{\rm s}(\overline{\mu }_R^2)}{Q^2 n_{f}}\frac{(1+\omega
)\overline{x}+(1-\omega
)x}{\omega \left [ (1+\omega )x+(1-\omega )\overline{x} \right ]}.
\end{equation}
 In deriving the hard-scattering amplitudes
$T_{1(2)}^g(x,Q^2,\omega, \mu _R^2)$
(they are the object of contradictory conclusions in the literature
\cite{ali2,pas}), the following projection
operator of the $\eta^{\prime}$-meson onto the two-gluon state has
been used
$$
P_{\mu \nu, ab}=i\frac{\delta_{ab}}{\sqrt{6} \sqrt{n_f}}
\frac{\epsilon_{\mu \nu \gamma \delta}q_1^{\gamma}q_2^{\delta}}
{\omega Q^2}.
$$
Note that in the limit $\omega = 0$ the difference $T_1^g - T_2^g$,
being the physical hard-scattering amplitude, is singularity-free
(see Eq.\ (\ref{eq:4.27}) below).
In the expressions above, $\alpha_{\rm s}(\mu^2)$ is the QCD
coupling constant in the two-loop approximation defined by
\begin{equation}
\label{eq:2.23}\alpha_{\rm s}(\mu ^2)
=\frac{4\pi }{\beta _0\ln (\mu ^2/\Lambda
^2)}
\left[ 1-\frac{2\beta _1}{\beta _0^2}\frac{\ln \ln (\mu ^2/\Lambda
^2)}{\ln
(\mu ^2/\Lambda ^2)}\right] ,
\end{equation}
with $\beta _0$ and $\beta _1$ being the one- and two-loop
coefficients of the QCD beta function:
\begin{equation}
\label{eq:2.24}\beta _0=11-\frac 23n_f,\,\,\,\beta _1
=51-\frac{19}3n_f.
\end{equation}
In Eqs.\ (\ref{eq:2.22}), (\ref{eq:2.23}), and (\ref{eq:2.24})
$\Lambda =0.3\,\,\rm{GeV}$
is the QCD scale parameter and $n_f$ is the number of active
quark flavors.

The physical quantity $F_{\eta ^{\prime }g^{*}g^{*}}(Q^2,\omega)$,
represented at sufficiently high $Q^2$ by the factorization formulas
(\ref{eq:2.16}), (\ref{eq:2.17}), is independent of the renormalization
scheme and the renormalization and factorization scales
$\mu _R^2\,(\overline{\mu }_R^2)$ and $\mu _F^2$.
Truncation of the perturbation series of $F_{\eta ^{\prime
}g^{*}g^{*}}(Q^2,\omega )$ at any finite order causes a residual
dependence on the scheme as well as on these scales.
At the leading order of pQCD, the hard-scattering amplitude does not
depend on the factorization scale $\mu _F^2$ but depends implicitly on
the renormalization scale $\mu_R^2$ through
$\alpha_{\rm s}(\mu_R^2)$.\footnote{Similar arguments hold for
the scale $\overline{\mu}_R^2$.}
An explicit dependence of the function $T_H$ on $\mu _R^2$ and
$\mu_F^2$ appears at $O(\alpha_{\rm s})$ due to QCD corrections.
As the scales $\mu_R^2$ and $\mu_F^2$ are independent of each other and
can be chosen separately, we adopt in this work the natural and widely
used ``default'' choice for the factorization scale $\mu _F^2=Q^2$ and
omit in what follows a dependence of the hard-scattering amplitude on
$\mu_F^2$ (see, Eqs.\ (\ref{eq:2.21}) and (\ref{eq:2.22})).
For a more detailed discussion of these questions, we refer the
interested reader to \cite{SSK00}.

In the standard HSA with a ``frozen'' coupling constant \cite{br1} one
sets the renormalization scale to be $\mu_R^2=\overline{\mu}_R^2=Q^2$,
simplifying the calculation of the vertex function
$F_{\eta^{\prime}g^{*}g^{*}}(Q^2,\omega)$ considerably.
In this approach the hard-scattering amplitudes $T_{1,2}^{q(g)}$ and
hence the quark and gluon components of the vertex function
$F_{\eta ^{\prime}g^{*}g^{*}}(Q^2,\omega)$ possess symmetry features
pertaining to the RC method.
Indeed, it is not difficult to see that
\begin{equation}
\label{eq:2.25}T_{1(2)}^q(\overline{x},Q^2,\omega
)=T_{2(1)}^q(x,Q^2,\omega),
\; \; T_{1(2)}^q(x,Q^2,-\omega)=T_{2(1)}^q(x,Q^2,\omega)
\end{equation}
and
\begin{equation}
\label{eq:2.26}T_{1(2)}^g(\overline{x},Q^2,\omega
)=T_{2(1)}^g(x,Q^2,\omega
),\;\; T_{1(2)}^g(x,Q^2,-\omega
)=-T_{2(1)}^g(x,Q^2,\omega).
\end{equation}
Using Eqs.\ (\ref{eq:2.10}) and (\ref{eq:2.25}), we then find
$$
F_{\eta ^{\prime }g^{*}g^{*}}^q(Q^2,\omega)=T_1^q(x,Q^2,\omega)
\otimes \phi ^q(x,Q^2)+(\omega \leftrightarrow -\omega )=
$$

\begin{equation}
\label{eq:2.27}=2T_1^q(x,Q^2,\omega)\otimes \phi ^q(x,Q^2).
\end{equation}

In similar manner, using Eqs.\ (\ref{eq:2.10}) and (\ref{eq:2.26}),
one can demonstrate that the following equalities hold
$$
F_{\eta ^{\prime }g^{*}g^{*}}^g(Q^2,\omega )=T_1^g(x,Q^2,\omega)
\otimes \phi ^g(x,Q^2)+(\omega\leftrightarrow -\omega )=
$$

\begin{equation}
\label{eq:2.28}=2T_1^g(x,Q^2,\omega )\otimes \phi ^g(x,Q^2).
\end{equation}
From Eqs.\ (\ref{eq:2.27}) and (\ref{eq:2.28}) it follows that
\begin{equation}
\label{eq:2.31}F_{\eta ^{\prime }g^{*}g^{*}}^q(Q^2,\omega )=F_{\eta
^{\prime }g^{*}g^{*}}^q(Q^2,-\omega ),\,\,\,F_{\eta ^{\prime
}g^{*}g^{*}}^g(Q^2,\omega )=F_{\eta ^{\prime
}g^{*}g^{*}}^g(Q^2,-\omega).
\end{equation}

In pQCD calculations higher-order corrections to various physical
processes, are, in general, large and in order to improve the
convergence of the corresponding perturbative series, different
methods have been proposed.
In exclusive processes - considering the pion electromagnetic form
factor as being a prominent example - the next-to-leading order
$O(\alpha_{\rm s})$ correction contains logarithmic terms
$\sim \ln(Q^2\overline{x}\overline{y}/\mu _R^2)$
(or $\sim \ln (Q^2xy/\mu _R^2)$) \cite{field,dittes},
which can be entirely or partly eliminated by a proper choice of the
renormalization scale $\mu _R^2\,\,\,(\overline{\mu }_R^2)$.
This can be achieved by taking as a renormalization scale
$\mu _R^2=Q^2\overline{x}\overline{y}\,\,\,\,[
\overline{\mu }_R^2=Q^2xy]$ or $\mu
_R^2=Q^2\overline{x}/2\,\,\,\,[\overline{\mu }_R^2=Q^2x/2]$.
The renormalization scale enters into the pQCD expression
not only via logarithmic terms, but also through the argument of the
running coupling $\alpha_{\rm s}(\mu _R^2)$.
In order to calculate the amplitude of an exclusive process, an
integration in corresponding integrals over the longitudinal momentum
fractions of the quarks and gluons in the involved hadrons has to be
carried out.
Thus, in the pion electromagnetic FF calculations, the integration over
the variables $x$ and $y$ has to be performed.
Traditionally, to avoid problems associated with singularities of
$\alpha_{\rm s}(\mu _R^2)$, one ``freezes'' the running coupling by
replacing $x$ and $y$, e.g., by their mean values
$x\rightarrow <x>=1/2,\;y\rightarrow <y>=1/2\;$ in the argument of
$\alpha_{\rm s}(\mu _R^2)$ and performs then the calculation with
$\alpha_{\rm s}(Q^2/4)$ \cite{field}.
Recently, the RC method for the calculation of various exclusive
processes with $\alpha_{\rm s}(\mu _R^2)$ was proposed.
Within this framework, one expresses the running coupling
$\alpha_{\rm s}( Q^2 x)$ in terms of $\alpha_{\rm s}(Q^2)$ by
employing the renormalization-group equation and performs the
integration over $x\,\,(y)$ using the principal value prescription.
This allows one to estimate power-suppressed corrections to the process
under consideration towards the endpoint region of phase space, thus
improving the agreement of QCD theoretical predictions with
experimental data.
An alternative option to calculate power-behaved contributions with
their coefficients would be to use the analytic perturbation theory
\cite{shir} along the lines proposed in \cite{stef,stef1}.
Next-to-leading order corrections are known for only a few exclusive
processes \cite{field,dittes,chase2}.
In general, the renormalization scale $\mu _R^2$ may be taken equal or
proportional to the square of the four-momentum $q^2$ of the virtual
parton(s) mediating the strong interaction in corresponding
leading-order Feynman diagrams.

 For the $\eta ^{\prime }$-meson on-shell gluon transition (i.e., for
$\omega =\pm 1$), we adopt in this paper the choice

\begin{equation}
\label{eq:2.32}\mu _R^2=Q^2x,\,\,\,\,\overline{\mu }_R^2
=Q^2\overline{x}.
\end{equation}
The renormalization scales (\ref{eq:2.32}) are equal to the momentum
squared $\vert q^2 \vert$ of the virtual partons (gluon or quark) in
the corresponding Feynman diagrams.
In the general case ($\omega \neq \pm 1$), the absolute value of the
square of the four-momenta $q^2$ of the virtual partons depends on both
the total gluon virtuality $Q^2$ and the asymmetry parameter $\omega$.
However, for the sake of simplifying the calculations and to avoid
problems related to the appearance of two parameters $Q^2$ and $\omega$
in the argument of $\alpha_{\rm s}$, we shall use a renormalization
scale of the form given in Eq.\ (\ref{eq:2.32}).
This choice is justified from the physical point of view because those
parts of the scales $\sim Q^2 x, \; \sim Q^2 \overline{x}$ are exactly
responsible for the power corrections $\sim 1/Q^{2n},\,n=1,\,2,..$
to the form factor $Q^2F_{\eta ^{\prime }g^{*}g^{*}}(Q^2,\omega)$ which
we are going to compute.

The next problem then is how to generalize Eqs.\ (\ref{eq:2.21}) and
(\ref{eq:2.22}) in such a way so that the corresponding hard-scattering
amplitudes $T_{1(2)}^{q(g)}$ and the quark and gluon components of the
vertex function will obey Eqs.\ (\ref{eq:2.25}-\ref{eq:2.31}).
For this purpose, we symmetrize the functions
$T_{1(2)}^{q(g)}(x,Q^2,\omega, \mu_R^{2}(\overline{\mu}_R^{2}))$
by exchanging $\mu _R^2\leftrightarrow \overline{\mu }_R^2$
to obtain
$$
T_1^q(x,Q^2,\omega)=\frac 12\left[ T_1^q(x,Q^2,\omega , \mu
_R^2)+T_1^q(x,Q^2,\omega , \overline{\mu }_R^2)\right]
$$

$$
=-\frac{2\pi }{3Q^2}\frac{\alpha_{\rm s}(Q^2x)+\alpha_{\rm
s}(Q^2\overline{x})}{
(1+\omega )x+(1-\omega )\overline{x}},
$$

\begin{equation}
\label{eq:2.33}T_2^q(x,Q^2,\omega)=\frac 12\left[ T_2^q(x,Q^2,\omega
,\overline{\mu }_R^2)+T_2^q(x,Q^2,\omega, \mu _R^2)\right]
\end{equation}

$$
=-\frac{2\pi }{3Q^2}\frac{\alpha_{\rm s}(Q^2x)+\alpha_{\rm
s}(Q^2\overline{x})}{
(1+\omega )\overline{x}+(1-\omega )x}
$$
and

$$
T_1^g(x,Q^2,\omega)=\frac 12\left[ T_1^g(x,Q^2,\omega, \mu
_R^2)+T_1^g(x,Q^2,\omega, \overline{\mu }_R^2)\right]
$$

$$
=\frac{\pi }{2Q^2n_f}\left[ \alpha_{\rm s}(Q^2x)+\alpha_{\rm
s}(Q^2\overline{x})\right]
\frac{(1+\omega )x+(1-\omega )\overline{x}}{\omega \left [
(1+\omega )\overline{x}+(1-\omega )x \right ] },
$$

$$
T_2^g(x,Q^2,\omega )=\frac 12\left[ T_2^g(x,Q^2,\omega ,
\overline{\mu }_R^2)+T_2^g(x,Q^2,\omega, \mu _R^2)\right]
$$

\begin{equation}
\label{eq:2.34}
=\frac{\pi }{2Q^2n_f}\left[ \alpha_{\rm s}(Q^2x)+\alpha_{\rm
s}(Q^2\overline{x}
)\right] \frac{(1+\omega )\overline{x}+(1-\omega )x
}{\omega \left [
(1+\omega )x+(1-\omega )\overline{x} \right ] }.
\end{equation}
In the ``frozen'' coupling-constant approximation
$Q^2x,\,\,Q^2 \overline{x}\rightarrow Q^2$,
Eqs.\ (\ref{eq:2.33}) and (\ref{eq:2.34})
coincide with expressions (\ref{eq:2.21}) and (\ref{eq:2.22}).
It is also not difficult to verify that the hard-scattering amplitudes
$T_{1(2)}^{q(g)}(x,Q^2,\omega )$
(\ref{eq:2.33}), (\ref{eq:2.34}) satisfy Eqs.\ (\ref{eq:2.25}) and
(\ref{eq:2.26}).
As a result, Eqs.\ (\ref{eq:2.27})-(\ref{eq:2.31}) remain valid also
within the RC method.

\subsection{Quark and gluon components of the $\mathbf{\eta^{\prime}}$
            meson distribution amplitude}
The important input information needed for the computation of the form
factor $F_{\eta ^{\prime }g^{*}g^{*}}(Q^2,\omega )$ are the DA's of the
quark and gluon components of the $\eta^{\prime}$ meson, namely, the
functions $\phi^q(x,Q^2)$ and $\phi ^g(x,Q^2)$.
In general, a meson DA is a function containing all nonperturbative,
long-distance effects, which cannot be calculated by employing
perturbative QCD tools.
Its dependence on $x$ (or its shape) has to be deduced from
experimental data or found using some nonperturbative methods, for
example, QCD sum-rules (see, e.g., \cite{BMS2001}).
In contrast, the evolution of the DA with the factorization scale $Q^2$
is governed by pQCD.

The evolution equation for the DA of a flavor singlet pseudoscalar
meson was derived and solved in Refs.\ \cite{ohrndorf,shif,ba}.
It turned out that due to mixing of the quark-antiquark and two-gluon
components of the meson DA, the evolution equation has a
$\left( 2\times 2\right) $ matrix form.
The anomalous dimensions matrix, which enters into the evolution
equation at the one-loop order, has the following components
\cite{ohrndorf, shif, ba} (see also Ref.\ \cite{bel})
$$
\gamma _{qq}^n=C_F\left[ 3+\frac 2{(n+1)(n+2)}-4\sum_{j=1}^{n+1}\frac
1j\right] ,\,\,\,\gamma _{gg}^n=N_c\left[ \frac{\beta _0}{N_c}+\frac
8{(n+1)(n+2)}-4\sum_{j=1}^{n+1}\frac 1j\right] ,
$$

\begin{equation}
\label{eq:2.35}\gamma _{qg}^n=\frac{12n_f}{(n+1)(n+2)},\,\,\,\,\,\gamma
_{gq}^n=C_F\frac{n(n+3)}{3(n+1)(n+2)},
\end{equation}
where $N_c=3$ and $C_F=4/3$ is the group theoretical factor for
$SU_c(3)$.
The anomalous dimensions matrix has the eigenvalues
\begin{equation}
\label{eq:2.36}\gamma _{\pm }^n=\frac 12
\left[ \gamma _{qq}^n+\gamma _{gg}^n\pm
\sqrt{(\gamma _{qq}^n-\gamma _{gg}^n)^2+4\gamma _{qg}^n\gamma _{gq}^n}
\right] .
\end{equation}
The solutions of the evolution equation for the quark and gluon
components of the DA has, in general, the form

\begin{equation}
\label{eq:2.37}\phi ^q(x,Q^2)=6Cx\overline{x}
\left\{ 1+\sum_{k=2,4..}^\infty
\left[ B_n^q\left( \frac{\alpha_{\rm s}(\mu _0^2)}{\alpha_{\rm
s}(Q^2)}\right) ^{\frac{
\gamma _{+}^n}{\beta _0}}+\rho _n^gB_n^g\left( \frac{\alpha_{\rm s}(\mu
_0^2)}{
\alpha_{\rm s}(Q^2)}\right) ^{\frac{\gamma _{-}^n}{\beta _0}}\right]
C_n^{3/2}(x-
\overline{x})\right\}
\end{equation}
and

\begin{equation}
\label{eq:2.38}\phi ^g(x,Q^2)=Cx\overline{x}\sum_{k=2,4..}^\infty
\left[ \rho
_n^qB_n^q\left( \frac{\alpha_{\rm s}(\mu _0^2)}{\alpha_{\rm
s}(Q^2)}\right) ^{\frac{
\gamma _{+}^n}{\beta _0}}+B_n^g\left( \frac{\alpha_{\rm s}(\mu
_0^2)}{\alpha_{\rm s}(Q^2)}\right) ^{\frac{\gamma _{-}^n}{\beta
_0}}\right] C_{n-1}^{5/2}(x-
\overline{x})
\end{equation}
with the constant $C$ being defined as
$$
C=\sqrt{2}f_q\sin \phi_p +f_s\cos \phi_p.
$$
In Eqs.\ (\ref{eq:2.37}) and (\ref{eq:2.38}) $C_n^{3/2}(z)$ and
$C_n^{5/2}(z)$ are Gegenbauer polynomials, calculable using the
recurrence formula \cite{bat1}

\begin{equation}
\label{eq:2.39}(n+1)C_{n+1}^\nu (z)=2(n+\nu )zC_n^\nu (z)-(n+2\nu
-1)C_{n-1}^\nu
(z),\,\,\,
\end{equation}

$$
C_0^\nu (z)=1,\;\;\;\;C_1^\nu (z)=2\nu z.
$$
The parameters $\rho_n^q$ and $\rho_n^g$ are determined in terms of the
anomalous dimensions matrix elements
\begin{equation}
\label{eq:2.40}
\rho _n^q=6\frac{\gamma _{+}^n-\gamma _{qq}^n}{\gamma _{gq}^n}
,\,\,\,\,\rho _n^g=\frac 16\frac{\gamma _{gq}^n}{\gamma _{-}^n-\gamma
_{qq}^n
}.
\end{equation}
In Eqs.\ (\ref{eq:2.37}) and (\ref{eq:2.38}) $\mu _0=1\,\,\rm{GeV}$
is the normalization point, at which the free input parameters
$B_n^q,\,\,B_n^g$ in the DA's $\phi^q(x,Q^2)$
and $\phi ^g(x,Q^2)$ have to be fixed.
Exactly these parameters determine the shape of the DA's.

For all $n\geq 2$, the eigenvalues $\gamma _{\pm }^n<0$ and their
absolute values increase with $n$.
Hence, in the asymptotic limit $Q^2\rightarrow \infty $, one has
$$
\left( \frac{\alpha_{\rm s}(\mu _0^2)}{\alpha_{\rm s}(Q^2)}\right)
^{\frac{\gamma
_{+}^n(\gamma _{-}^n)}{\beta _0}}
\sim \ln \left( Q^2/\Lambda ^2\right)^{
\frac{\gamma _{+}^n(\gamma _{-}^n)}{\beta _0}}\rightarrow 0
$$
and therefore only the quark component of the $\eta ^{\prime }$ meson
DA survives, evolving to its asymptotic form, whereas the gluon DA in
this limit vanishes,
\begin{equation}
\label{eq:2.41}
\phi ^q(x,Q^2)\stackrel{Q^2\rightarrow \infty }{\longrightarrow
}
\,\,6Cx\overline{x},\,\,\,\,\,\,\phi ^g(x,Q^2)\stackrel{Q^2\rightarrow
\infty
}{\longrightarrow }\,\,\,0.
\end{equation}

In this work, we shall use the $\eta^{\prime}$-meson DA that contains
only the first non-asymptotic terms.
In other words, we suggest that in Eqs.\ (\ref{eq:2.37}) and
(\ref{eq:2.38})
$B_2^q\neq 0,\,\,B_2^g\neq 0$ and $B_n^q=B_n^g=0$ for all $n\geq
4$.
The numerical values of the relevant parameters for $n_f=3$ are
$$
\gamma _{qq}^2=-\frac{50}9,\,\,\,\,\,\gamma _{gg}^2=-11,
\,\;\;\;\,\,\gamma_{gq}^2=\,\frac{10}{27},\,\,\,\,\gamma _{qg}^2=3,
$$
\begin{equation}
\label{eq:2.42}\gamma _{+}^2\simeq -\frac{48}9,\,\,\,\gamma _{-}^2
\simeq
-\frac{
101}9\,,\,\,\,\rho_2^q
\simeq \frac{16}5,\,\,\,\,\,\rho _2^g\simeq -\frac
1{90}.
\end{equation}
Taking into account Eq.\ (\ref{eq:2.42}) and the expressions for the
required Gegenbauer polynomials
$$
C_2^{3/2}(x-\overline{x})=\frac 32\left[ 5(x-\overline{x})^2-1\right]
=6\left( 1-5x\overline{x}\right) ,
\,\,\,\,C_1^{5/2}(x-\overline{x})\,=5(x-\overline{x}),
$$
we finally recast the $\eta^{\prime}$ meson quark and gluon DA's into
the more suitable (for our purposes) forms
\begin{equation}
\label{eq:2.43}\phi ^q(x,Q^2)=6Cx\overline{x}\left[
1+A(Q^2)-5A(Q^2)x\overline{x}
\right] ,\,\,\,\,\,\,\,\phi ^g(x,Q^2)
=Cx\overline{x}(x-\overline{x})B(Q^2).
\end{equation}
In these expressions the functions $A(Q^2)$ and $B(Q^2)$ are defined by
$$
A(Q^2)=6B_2^q\left( \frac{\alpha_{\rm s}(Q^2)}{\alpha_{\rm s}(\mu
_0^2)}\right) ^{
\frac{48}{81}}-\frac{B_2^g}{15}\left( \frac{\alpha_{\rm
s}(Q^2)}{\alpha_{\rm s}(\mu
_0^2)}\right) ^{\frac{101}{81}},
$$
\begin{equation}
\label{eq:2.44}
B(Q^2)=16B_2^q\left( \frac{\alpha_{\rm s}(Q^2)}{\alpha_{\rm
s}(\mu _0^2)}
\right) ^{\frac{48}{81}}+5B_2^g\left( \frac{\alpha_{\rm
s}(Q^2)}{\alpha_{\rm s}(\mu
_0^2)}\right) ^{\frac{101}{81}}.
\end{equation}
 The $\eta^{\prime}$ meson quark and gluon DA's for $n_f=4$
are given in the Appendix. Equations (\ref{eq:2.43}), (\ref{eq:2.44}),
and (\ref{eq:1app}) contain all necessary information about the
singlet part of the $\eta^{\prime}$ meson DA.

\section{Running coupling method and power suppressed corrections}
\setcounter{equation}0
To compute the $\eta^{\prime}$-meson virtual-gluon transition form
factor $F_{\eta^{\prime}g^{*}g^{*}}(Q^2,\omega )$, we have to perform
in Eqs.\ (\ref{eq:2.16}) and (\ref{eq:2.17}) the integration over $x$
by inserting the explicit expressions for the hard-scattering
amplitudes $T_{1,2}^{q(g)}(x,Q^2,\omega )$ and the $\eta^{\prime}$-meson
quark and gluon DA's, and retain the $x$-dependence of the coupling
$\alpha_{\rm s}(Q^2x)$ [$\alpha_{\rm s}(Q^2 \overline{x})$].
Such calculations lead to divergent integrals because the running
coupling $\alpha_{\rm s}(Q^2x)$ [$\alpha_{\rm s}(Q^2 \overline{x})$]
suffers from an infrared singularity in the limit
$x\rightarrow 0$ [$x\rightarrow 1$].
This means that in order to carry out computations with the running
coupling, some procedure for its regularization in the end-point
$x\rightarrow 0,\,1$ regions has to be adopted.

As a first step in this direction, we express the running coupling
$\alpha_{\rm s}(Q^2 x)$ in terms of $\alpha_{\rm s}(Q^2)$,
employing the renormalization-group equation
\begin{equation}
\label{eq:3.1}\frac{\partial \alpha_{\rm s}(Q^2 x)}{\partial \ln
x }=-
\frac{\beta _0}{4\pi }\left[ \alpha_{\rm s}(Q^2 x )\right]
^2-\frac{\beta _1
}{8\pi ^2}\left[ \alpha_{\rm s}(Q^2 x )\right] ^3.
\end{equation}
The solution of Eq.\ (\ref{eq:3.1}), obtained by keeping the leading
$(\alpha_{\rm
s}\ln x )^k$ and next-to-leading
$\alpha_{\rm s}(\alpha_{\rm s}\ln x)^{k-1}$
powers of $\ln x$, reads \cite{st}
\begin{equation}
\label{eq:3.2}\alpha_{\rm s}(Q^2 x )\simeq \frac{\alpha_{\rm
s}(Q^2)}{1+\ln x
/t}\left[ 1-\frac{\alpha_{\rm s}(Q^2)\beta _1}{2\pi \beta _0}\frac{\ln
[1+\ln
x /t]}{1+\ln x /t}\right] ,
\end{equation}
where $\alpha_{\rm s}(Q^2)$ is the one-loop QCD coupling constant and
$t=4\pi
/\beta _0\alpha_{\rm s}(Q^2)$.

Inserting Eq.\ (\ref{eq:3.2}) into the expressions for the
hard-scattering amplitudes and subsequently
$T_{1,2}^{q(g)}(x,Q^2,\omega)$ into Eqs.\ (\ref{eq:2.16}) and
(\ref{eq:2.17}), we obtain integrals which are still divergent, but
which now can be calculated using existing methods.
One of these methods, applied in Ref.\ \cite{ag2} for the calculation
of the electromagnetic pion form factor, allows us to find the
quantity under consideration as a perturbative series in
$\alpha_{\rm s}(Q^2)$ with factorially growing coefficients
$C_n\sim (n-1)!$.
Similar series with coefficients $C_n\sim (n-1)!$ may be found also
for the form factor
$Q^2F_{\eta^{\prime }g^{*}g^{*}}(Q^2,\omega)$, i.e.,
\begin{equation}
\label{eq:3.3}Q^2F_{\eta ^{\prime }g^{*}g^{*}}(Q^2,\omega )\sim
\sum_{n=1}^\infty
\left[ \frac{\alpha_{\rm s}(Q^2)}{4\pi}\right]^n\beta_0^{n-1}C_n.
\end{equation}
But a perturbative QCD series with factorially growing coefficients is
a signal for the IR renormalon nature of the divergences in
Eq.\ (\ref{eq:3.3}).
The convergence radius of the series (\ref{eq:3.3}) is zero and its
resummation should be performed by employing the Borel integral
technique.
First, one has to find the Borel transform
$B[Q^2F_{\eta ^{\prime }g^{*}g^{*}}](u)$ of the corresponding
series \cite{zakh}
\begin{equation}
\label{eq:3.4}B\left[ Q^2F_{\eta ^{\prime }g^{*}g^{*}}\right]
(u)=\sum_{n=1}^\infty \frac{u^{n-1}}{(n-1)!}C_n.
\end{equation}
Because the coefficients of the series (\ref{eq:3.3}) behave like
$C_n\sim (n-1)!$, the Borel transform (\ref{eq:3.4}) contains poles
located at the positive $u$ axis of the Borel plane.
In other words, the divergence of the series (\ref{eq:3.3}) owing to
Eq.\ (\ref{eq:3.4})  has been transformed into pole-like singularities
of the function $B[Q^2F_{\eta^{\prime}g^{*}g^{*}}](u)$.
These poles are the footprints of the IR renormalons.

Now in order to define the sum (\ref{eq:3.3}) or to find the resummed
expression for the form factor, one has to invert
$B[Q^2F_{\eta ^{\prime}g^{*}g^{*}}](u)$ to get
\begin{equation}
\label{eq:3.5}\left[ Q^2F_{\eta ^{\prime }g^{*}g^{*}}(Q^2,\omega
)\right]
^{res}\sim P.V.\int_0^\infty du\exp \left[ -\frac{4\pi }{\beta
_0\alpha_{\rm s}(Q^2)}u\right] B\left[ Q^2F_{\eta ^{\prime
}g^{*}g^{*}}\right] (u)
\end{equation}
and remove the IR renormalon divergences by the principal value
prescription.

The procedure described above, is straightforward but, at the same
time, tedious.
Fortunately, these intermediate operations can be bypassed by
introducing the inverse Laplace transforms of the functions in
Eq.\ (\ref{eq:3.2}) \cite{bat2}, viz.,

\begin{equation}
\label{eq:3.6}\frac 1{(t+z)^\nu }
=\frac 1{\Gamma (\nu )}\int_0^\infty du\exp
[-u(t+z)]u^{\nu -1},\,\,\,\,\,\,\,Re\nu >0
\end{equation}
and

\begin{equation}
\label{eq:3.7}\frac{\ln (t+z)}{(t+z)^2}=\int_0^\infty du\exp
[-u(t+z)](1-\gamma
_E-\ln u)u,\,\,\,\,\,\,\,
\end{equation}
where $\Gamma (z)$ is the Gamma function, $\gamma _E\simeq 0.577216$
is the Euler constant, and $z=\ln x$ [or $z=\ln \overline{x}$].

Then using Eqs.\ (\ref{eq:3.2}), (\ref{eq:3.6}) and (\ref{eq:3.7})
for $\alpha_{\rm s}(Q^2 x )$, we get \cite{ag1}

\begin{equation}
\label{eq:3.8}\alpha_{\rm s}(Q^2 x )
=\alpha_{\rm s}(Q^2)t\int_0^\infty
du\exp
(-ut) x^{-u}R(u,t)=\frac{4\pi }{\beta _0}\int_0^\infty du\exp
(-ut) x^{-u}R(u,t)
\end{equation}
with the function $R(u,t)$ defined as

\begin{equation}
\label{eq:3.9}R(u,t)
=1-\frac{2\beta _1}{\beta _0^2}u(1-\gamma _E-\ln t-\ln
u).
\end{equation}
Employing Eq.\ (\ref{eq:3.8}) for $\alpha_{\rm s}(Q^2 x )$ and
carrying out the integrations over $x$, we obtain (see, the next
Section) the quark and gluon components of the form factor
$F_{\eta^{\prime }g^{*}g^{*}}(Q^2,\omega)$ directly in the Borel
resummed form (\ref{eq:3.5}).

The resummed vertex function
$[Q^2F_{\eta ^{\prime}g^{*}g^{*}}(Q^2,\omega)]^{res}$
- except for the $\eta ^{\prime }$-meson on-shell-gluon transition
and the $\omega = 0$ case -
contains an infinite number of IR renormalon poles located at $u_0=n$,
where $n$ is a positive integer that, in general, is given as a sum of
a series, i.e.,
\begin{equation}
\label{eq:3.10}S(k_0,Q^2,\omega)\sim \sum_{n=k_0}^\infty \frac{4\pi
}{
\beta _0} P.V. \int_0^\infty \frac{e^{-ut}R(u,t)du}{n-u}C_n(\omega).
\end{equation}
One appreciates that all information on the process asymmetry parameter
$\omega $ is accumulated in the coefficient functions $C_n(\omega)$.

To avoid inessential technical details and to make the presentation as
transparent as possible, let us for the time being neglect the
non-leading term $\sim \alpha_{\rm s}^2$ in Eq.\ (\ref{eq:3.2}) and
make the replacement $R(u,t)\rightarrow 1$ in Eq.\ (\ref{eq:3.10}).
Then, the integrals in Eq.\ (\ref{eq:3.10}) take, after simple
manipulations, the form
\begin{equation}
\label{eq:3.11}\frac{4\pi }{\beta _0} P.V. \int_0^\infty
\frac{e^{-ut}du}{n-u}=\frac{
4\pi }{\beta _0}\frac{li(\lambda ^n)}{\lambda ^n},\;\;\;\lambda
=Q^2/\Lambda
^2
\end{equation}
with the logarithmic integral $li(\xi)$ defined by
\begin{equation}
\label{eq:3.12}li(\xi)=P.V.\int_0^{\xi}\frac{dt}{\ln t}.
\end{equation}
The function $li(\xi)$ is expressible in terms of the function $E_1(z)$
\cite{bat1}
\begin{equation}
\label{eq:3.13}li(\xi)=-E_1(-\ln \xi),
\end{equation}
where
$$
E_1(z)=\int_z^\infty e^{-t}t^{-1}dt.
$$
The latter expression can, for $\left| z\right| \rightarrow \infty $,
be expanded over inverse powers of $z$

\begin{equation}
\label{eq:3.14}E_1(z)=z^{-1}e^{-z}\left[
\sum_{m=0}^M\frac{m!}{(-z)^m}+O\left(
\left| z\right| ^{-(M+1)}\right) \right] .
\end{equation}
Using Eqs.\ (\ref{eq:3.13}) and (\ref{eq:3.14}), we find

\begin{equation}
\label{eq:3.15}
li(\xi)\simeq \frac {\xi}{\ln \xi}\sum_{m=0}^M\frac{m!}{(\ln \xi)^m}
,\,\,\,\,\,\,\,\,\,\,\,\,\,\frac{li({\xi}^n)}{{\xi}^n}
\simeq \frac 1{n\ln
\xi}\sum_{m=0}^M\frac{m!}{(n\ln \xi)^m}.
\end{equation}
It is not difficult to see that for $Q^2\gg \Lambda ^2$

\begin{equation}
\label{eq:3.16}\frac{4\pi}{\beta_0}\frac{li(\lambda^n)}{\lambda ^n}
\simeq
4\pi \sum_{m=1}^M\left[ \frac{\alpha_{\rm s}(Q^2)}{4\pi}\right]^m\beta
_0^{m-1}
\frac{(m-1)!}{n^m}.
\end{equation}
Choosing in Eq.\ (\ref{eq:3.16}) $M\rightarrow \infty $ and comparing
the obtained result with Eq.\ (\ref{eq:3.3}), we conclude that each
term in the sum (\ref{eq:3.10}) at fixed $n$ is the resummed Borel
expression of the series (\ref{eq:3.16}).

But we can look at the integrals in Eq.\ (\ref{eq:3.10}) from another
perspective.
Namely, it is easy to show that

\begin{equation}
\label{eq:3.17}
\frac{4\pi }{\beta _0}\int_0^\infty \frac{e^{-ut}du}{n-u}
=\int_0^1\alpha_{\rm s}(Q^2x)x^{n-1}dx=\frac 1nf_{2n}(Q),
\end{equation}
where $f_{2n}(Q)$ are the moment integrals

\begin{equation}
\label{eq:3.18}f_p(Q)
=\frac p{Q^p}\int_0^Qdkk^{p-1}\alpha_{\rm s}(k^2).
\end{equation}
The integrals $f_p(Q)$ can be calculated using the IR matching scheme
\cite{web}
\begin{equation}
\label{eq:3.19}f_p(Q)=
\left( \frac {\mu_{I}} {Q}\right) ^pf_p(\mu_I )+
\alpha_{\rm s}(Q^2)\sum_{n=0}^N
\left[ \frac{\beta _0}{2\pi p}\alpha_{\rm
s}(Q^2)\right]
^n\left[ n!-\Gamma \left( n+1,\,\,p\ln (Q/\mu_I )\right) \right] ,
\end{equation}
where $\mu_{I}$ is the infrared matching scale and $\Gamma (n+1,z)$ is
the incomplete Gamma function

\begin{equation}
\label{eq:3.20}\Gamma \left( n+1,\,z\right) =\int_z^\infty e^{-t}t^ndt.
\end{equation}

In Eq.\ (\ref{eq:3.19}) $f_p(\mu_{I} )$ are phenomenological parameters,
which represent the weighted average of $\alpha_{\rm s}(k^2)$ over the
IR region $0<k<\mu_I $ and act at the same time as infrared regulators
of the r.h.s. of Eq.\ (\ref{eq:3.17}).
The first term on the r.h.s. of Eq.\ (\ref{eq:3.19}) is a
power-suppressed contribution to $f_p(Q)$, which cannot be calculated
within pQCD, whereas the second term is the perturbatively calculable
part of the function $f_p(Q)$.
Stated differently, the infrared matching scheme allows one to estimate
power corrections to the moment integrals by explicitly dissecting them
from the full expression and introducing new nonperturbative parameters
$f_p(\mu_{I} )$.
The values of these parameters have to be extracted form experimental
data.

In order to check the accuracy of both the RC method (IR renormalon
calculus) and the infrared matching scheme, a Landau-pole free model
for $\alpha_{\rm s}(Q^2)$ was introduced in Ref.\ \cite{web} yielding
\begin{equation}
\label{eq:3.21}\alpha_{\rm s}(Q^2)=\frac{4\pi }{\beta _0}\left[ \frac 1
{\ln\lambda}+125\frac{
1+4\lambda }{(1-\lambda )(4+\lambda )^4}\right].
\end{equation}

The model expression (\ref{eq:3.21}) was used for the exact calculation
of the moment integrals $f_p^{\rm{exact}}(Q)$.
The IR renormalon and the IR matching scheme results were compared with
$f_p^{\rm{exact}}(Q)$ concluding that for $N\geq 2$ the IR matching
scheme leads to a good agreement with the exact values of $f_p(Q)$.
As regards the calculation of the integrals $f_p(Q)$, when employing
the RC method with the principal value prescription, a good agreement
with $f_2^{\rm{exact}}(Q)$ was found for $p=2$, while for $p=1$ a small
deviation from $f_1^{\rm{exact}}(Q)$ in the region of $Q^2\sim $ a few
$\rm{GeV}^2$ was observed (the lowest moment integral entering into our
expressions is $f_2(Q)$).
But even for $p=1$, the agreement with $f_1^{\rm{exact}}(Q)$ within the
error bars is quite satisfactory in view of the uncertainties produced
by the principal value prescription itself.
Therefore, we can conclude that the RC method and the IR matching
scheme lead for the moment integrals $f_p(Q)$ to approximately similar
numerical results.
It is remarkable that in both cases an enhancement of the IR renormalon
and matching-scheme results relative to the leading-order perturbative
prediction was found.

After presenting our analysis, it becomes evident that the Borel
resummation technique enables us to estimate with a rather good
accuracy power-behaved corrections to the integrals in
Eq.\ (\ref{eq:3.10}) and hence to the vertex function
$F_{\eta^{\prime}g^{*}g^{*}}(Q^2,\omega)$.
Of course, one can employ the IR matching scheme and
Eq.\ (\ref{eq:3.19}) to determine power corrections to the form
factor in explicit form.
But as we have noted above, in general, the form factor
$Q^2F_{\eta^{\prime }g^{*}g^{*}}(Q^2,\omega)$
contains an infinite number of IR renormalon poles or stated
equivalently, an infinite number of $f_{2n}(Q)$ is required to
calculate sums like the one presented in Eq.\ (\ref{eq:3.10}).
In real numerical computations, in order to reach a good approximation
of the series (\ref{eq:3.10}), at least a sum of the first $15\div 20$
terms is needed, meaning the involvement of
$f_{2n}(\mu_{I} )$ ($n=15\div 20$) nonperturbative parameters.
But from the experimental data, only the first few moments
$f_p(\mu_{I})$ could be extracted \cite{das1,mov,H1, NMC}.
Values for $f_1(2\,\,\,\rm{GeV})$ (in the literature the notation
$f_1(\mu_I)=\alpha_0(\mu_I)$ is widely used) range from
$\alpha_0(2\,\,\rm{GeV})=0.435 \pm 0.021$ \cite{das1} and
$\alpha_0(2\,\, {\rm GeV})=0.513^{+0.066}_{-0.045}$ \cite{mov}
to $\alpha_0(2\,\,{\rm GeV})=0.597^{+0.009}_{-0.010}$ \cite{H1}.
Here, we write down only sample results, obtained in these works, in
order to demonstrate the experimental situation with
$\alpha_0(2\,\,\rm{GeV})=f_1(2\,\,\rm{GeV})$.
For $f_2(2\,\,\,\rm{GeV})$
the value $f_2(2\,\,\rm{GeV})\simeq 0.5$ was deduced \cite{das2,web}
from the data on structure functions \cite{NMC}.

In the framework of the RC method, we estimate the same power
corrections to a physical quantity, but here we do not need additional
information on $f_p(\mu_{I})$.
Moreover, this method gives us the possibility to calculate values of
the nonperturbative parameters $f_p(\mu_{I} )$ by means of the formula
\begin{equation}
\label{eq:3.22}f_{2n}(\mu_{I} )=n\frac{4\pi }{\beta _0}\frac{li\left(
\widetilde{
\lambda }^n\right) }{\widetilde{\lambda }^n},\,\,\,\,\,\,\,\,
\widetilde{
\lambda }=\mu_{I} ^2/\Lambda ^2
\end{equation}
that can be easily found by comparing Eqs.\ (\ref{eq:3.11}) and
(\ref{eq:3.17}).
The calculated values of the first few even moments
($n_f=3,\,\,\,\,\Lambda =0.3\,\,\,\rm{GeV}$) are
\begin{equation}
\label{eq:3.23}
f_2(2\,\,\rm{GeV})\simeq 0.535,\,\,\,\,f_4(2\,\,\rm{GeV})
\simeq
0.45,\,\,f_6(2\,\,\rm{GeV})\simeq 0.41.
\end{equation}
To compare our predictions with those computed via the model
$\alpha_{\rm s}(Q^2)$ (Eq.\ (\ref{eq:3.21})), we choose $n_f=3$ and
$\Lambda =0.25\,\,\rm{GeV}$.
The corresponding values are shown below
\begin{equation}
\label{3.24}
\begin{array}{c}
\,\,f_2^{\rm{RC}}(2\,\,\rm{GeV})\simeq
0.479,\,\,\,\,\,\,\,f_4^{\rm{RC}}(2\,\,\rm{GeV})\simeq
0.393, \\
\,f_2^{\rm{mod}}(2\,\,\rm{GeV})\simeq 0.450,\,\,\,\,\,\,\,
f_4^{\rm{mod}}(2\,\,\rm{GeV})\simeq
0.388.
\end{array}
\end{equation}
One observes that the parameters $f_p(\mu_I )$ of the RC method are in
agreement with both the experimental results and the model
calculations.

We have noted above that the principal value prescription, adopted in
this work in order to regularize divergent integrals (see,
Eqs.\ (\ref{eq:3.5}), (\ref{eq:3.10})) generates power-suppressed
(higher-twist) uncertainties
\begin{equation}
\label{eq:3.25}
\sim \sum_q N_q \frac {\Phi_q(Q^2)}{Q^{2q}},
\end{equation}
where $\{\Phi_q(Q^2)\}$ are calculable functions entirely fixed by the
residues of the Borel transform $B[Q^2F_{Mg^{*}g^{*}}](u)$ at the
poles $q=u_0$ and $\{N_q\}$ are arbitrary constants to be fixed
from experimental data.
In the case of the $\eta^{\prime}\gamma$ electromagnetic transition
form factor $Q^2F_{\eta^{\prime}\gamma}(Q^2)$, the uncertainties
(\ref{eq:3.25}) were estimated in Ref.\ \cite{ag1}.
For the $\eta^{\prime}$ meson asymptotic DA and for the parameters
$N_1=N_2=1$ and $N_1=N_2=-1$, it was found that they do not exceed
$\pm 15 \%$ of the $\eta^{\prime}\gamma$ form factor.
Because the RC method allows one to evaluate power corrections to a
physical quantity, in general, and for the $\eta^{\prime} \gamma$
transition form factor it has already provided a good agreement with
the CLEO data \cite{ag1}, one has to introduce the constraint
$\left|N_{1,2} \right|\ll 1$ in order to retain this agreement.
In the present work we shall estimate in our numerical analysis the
higher-twist uncertainties produced by the principal value
prescription by choosing the constants $\{ N_q \}=\pm 1$.
It will be demonstrated that the $\pm 15 \%$ bounds are valid also
for the $\eta^{\prime}g,\; \eta^{\prime}g^{*}$ transition FF's.
Inclusion of higher-twist corrections, arising from the two particle
higher-twist and higher Fock states DA's of the $\eta^{\prime}$ meson,
may, in principle, further improve this  scheme.
These refinements are, however, beyond the scope of the present
investigation.
\thinspace

\section{The form factor
         $\mathbf{F_{\eta^{\prime}g^{*}g^{*}}(Q^2,\omega)}$
         within the RC method}
\setcounter{equation}0
In this section we calculate the quark and gluon components of the
vertex function $F_{\eta^{\prime}g^{*}g^{*}}(Q^2, \omega)$ within
the RC method. We also present our results for
$F_{\eta^{\prime}gg^{*}}(Q^2, \omega=\pm 1)$,
$F_{\eta^{\prime}g^{*}g^{*}}(Q^2,\omega=0)$ and the asymptotic limit for
the form factor.

\subsection{Quark component
            $\mathbf{F_{\eta^{\prime}g^{*}g^{*}}^q(Q^2, \omega)}$
            of the vertex function}

To calculate the quark component of the vertex function
$F_{\eta^{\prime }g^{*}g^{*}}^q(Q^2,\omega)$, we use
Eq.\ (\ref{eq:2.27}) for
$F_{\eta ^{\prime }g^{*}g^{*}}^q(Q^2,\omega)$,
Eq.\ (\ref{eq:2.33}) for the hard scattering amplitude
$T_1^q(x,Q^2,\omega)$, and the expression for
$\alpha_{\rm s}(Q^2 x)$ given in Eq.\ (\ref{eq:3.8}).
Employing the quark component of the $\eta ^{\prime }$-meson DA,
$\phi ^q(x,Q^2)$, we get
\footnote{In what follows integrals over the variable $u$
have to be understood in the sense of the Cauchy principal value.}

$$
Q^2F_{\eta ^{\prime }g^{*}g^{*}}^q(Q^2,\omega)=-\frac{32\pi
^2C[1+A(Q^2)]}{\beta _0}\int_0^\infty due^{-ut}R(u,t)
$$

$$
\times\left[\int_0^1\frac{x^{1-u}\overline{x}dx}{(1+\omega )x+(1-\omega
)
\overline{x}}+\int_0^1\frac{x\overline{x}^{1-u}dx}{
(1+\omega )x+(1-\omega )\overline{x}}\right]
$$

$$
+\frac{160\pi ^2CA(Q^2)}{\beta _0}\int_0^\infty due^{-ut}R(u,t)\left[
\int_0^1 \frac{x^{2-u}\overline{x}^2dx}{(1+\omega )x+(1-\omega
)\overline{x}
 }\right.
$$

\begin{equation}
\label{eq:4.1}\left. +\int_0^1\frac{x^2\overline{x}^{2-u}dx}{(1+\omega
)x+(1-\omega )\overline{x}}\right] .
\end{equation}

Using the fact that the integral \cite{pr1}
\begin{equation}
\label{eq:4.2}
\int_0^1 x^{\alpha -1}\overline{x}^{\beta -1}(1-xr)^{\gamma}dx
=B(\alpha ,\beta )_2F_1\left(
-\gamma, \alpha; \alpha +\beta ;r \right )
\end{equation}
is expressible in terms of the Gauss hypergeometric function \cite{pr2}

\begin{equation}
\label{eq:4.3}_2F_1(a,b;c;z)=\sum_{k=0}^\infty
\frac{(a)_k(b)_k}{(c)_k}\frac{
z^k}{k!},
\end{equation}
we obtain

$$
Q^2F_{\eta^{\prime}g^{*}g^{*}}^q(Q^2,\omega)=
-\frac{32 \pi^2C[1+A(Q^2)]}{\beta_0(1+\omega)}
\int_{0}^{\infty}due^{-ut}R(u,t)B(2-u,2)
$$

$$
\times \left [_2F_1\left (1,2;4-u;\frac{2\omega}{1+\omega} \right)+
{}_2F_1 \left (1,2-u;4-u; \frac{2\omega}{1+\omega} \right )\right ]
$$

$$
+\frac{160\pi^2CA(Q^2)}{\beta_0(1+\omega)}
\int_{0}^{\infty}due^{-ut}R(u,t)
B(3-u,3)
\left[{}_2F_1 \left (1,3;6-u; \frac{2\omega}{1+\omega}\right) \right .
$$

\begin{equation}
\label{eq:4.4}
\left . +{}_2F_1 \left ( 1,3-u;6-u;
\frac{2\omega}{1+\omega}\right )\right].
\end{equation}

In Eqs.\ (\ref{eq:4.2}) and (\ref{eq:4.3}), $B(x,y)$ and $(a)_n$ are
the Beta function and Pochhammer symbols, respectively, defined in
terms of the Gamma function $\Gamma (z)$
$$
B(x,y)=\frac{\Gamma (x)\Gamma (y)}{\Gamma (x+y)},\;\; (a)_n
=\frac{\Gamma(a+n)}{\Gamma (a)}.
$$

From Eq.\ (\ref{eq:4.4}) a simple expression for
the $\eta^{\prime}$-meson on-shell-gluon transition form factor
$F_{\eta^{\prime}gg^{*}}^q(Q^2, \omega = \pm 1)$ can be found

$$
Q^2F_{\eta ^{\prime }gg^{*}}^q(Q^2,\omega = \pm 1)=-\frac{16\pi
^2C[1+A(Q^2)]}{\beta _0}\int_0^\infty due^{-ut}R(u,t)\left[
B(1,2-u)\right.
$$
\begin{equation}
\label{eq:4.7}
\left. +B(2,1-u)\right]+\frac{80\pi ^2CA(Q^2)}{\beta _0}
\int_0^\infty due^{-ut}R(u,t)\left[ B(3,2-u)+B(2,3-u)\right] ,
\end{equation}
where we have used the equality \cite{pr2}

\begin{equation}
\label{eq:4.8}\,_2F_1(a,b;c;1)
=\frac{\Gamma (c)\Gamma (c-a-b)}{\Gamma
(c-a)\Gamma (c-b)} .
\end{equation}
It is worth noting that Eq.\ (\ref{eq:4.7}) can be obtained from
Eq.\ (\ref{eq:2.27}) by employing the
$\omega\to\pm 1$ limits of the hard scattering amplitudes
(\ref{eq:2.33})

\begin{equation}
\label{eq:4.9}T_1^q(x,Q^2,\omega =\pm 1)+T_2^q(x,Q^2,\omega=\pm
1)
=-\frac{\pi}{3Q^2}\left[\alpha_{\rm s}(Q^2x)+\alpha_{\rm
s}(Q^2\overline{x})\right ]
\left[\frac 1 {x}+\frac 1 {\overline{x}}\right ] .
\end{equation}
In the case of gluons with equal virtualities $Q_1^2=Q_2^2$ (and
hence in the $\omega=0$ case), the form factor can be found
employing the expression for the hard-scattering amplitude

\begin{equation}
\label{eq:4.10}
T_1^q(x,Q^2,\omega=0)+T_2^q(x,Q^2,\omega=0)=
-\frac{4\pi}{3Q^2}\left[\alpha_{\rm s}(Q^2x)
+\alpha_{\rm s}(Q^2\overline{x})\right ].
\end{equation}
Calculation leads to the simple expression

$$
Q^2F_{\eta^{\prime}g^{*}g^{*}}^q(Q^2,\omega=0)
=-\frac{64\pi^2C(1+A(Q^2))}{\beta_0}
\int_0^{\infty}due^{-ut}R(u,t)B(2-u,2)
$$
\begin{equation}
\label{eq:4.11}
+\frac{320\pi^2CA(Q^2)}{\beta_0}
\int_0^{\infty}due^{-ut}R(u,t)B(3-u,3).
\end{equation}
Let us note that Eq.\ (\ref{eq:4.11}) can be deduced from
Eq.\ (\ref{eq:4.4}) in the limit $\omega \to 0$
by taking into account that \cite{pr2}
$$
{}_2F_1(a,b;c;0)=1.
$$

The next important problem to be solved within the framework of the
RC method, is to reveal the IR renormalon poles in
Eqs.\ (\ref{eq:4.4}), (\ref{eq:4.7}) and
(\ref{eq:4.11}) because without such a clarification all these
expressions would have merely a formal character.

We start from the simple case, i.e., from Eq.\ (\ref{eq:4.7}), and get
$$
B(1,2-u)+B(2,1-u)=\frac{\Gamma (1)\Gamma (2-u)}{\Gamma (3-u)}
+\frac{\Gamma
(2)\Gamma (1-u)}{\Gamma (3-u)}
$$
$$
=\frac 1{2-u}+\frac 1{(1-u)(2-u)}=\frac 1{1-u}
$$
and
$$
B(3,2-u)+B(2,3-u)=\frac{\Gamma (3)\Gamma (2-u)}{\Gamma (5-u)}
+\frac{\Gamma
(2)\Gamma (3-u)}{\Gamma (5-u)}
$$
$$
=\frac 2{(2-u)(3-u)(4-u)}+\frac 1{(3-u)(4-u)}
=\frac 1{2-u}-\frac 1{3-u}.
$$
In deriving these expressions we used the following property of the
Gamma function
$$
z\Gamma (z)=\Gamma (z+1).
$$
Hence we have
$$
Q^2F_{\eta ^{\prime }gg^{*}}^q(Q^2,\omega =\pm 1)=-\frac{16\pi
^2C[1+A(Q^2)]}{\beta _0}\int_0^\infty e^{-ut}R(u,t)\frac{du}{1-u}
$$
\begin{equation}
\label{eq:4.12}+\frac{80\pi ^2CA(Q^2)}{\beta _0}\int_0^\infty
e^{-ut}R(u,t)\left( \frac 1{2-u}-\frac 1{3-u}\right) du.
\end{equation}
As one sees, the structure of the IR renormalon poles is very simple;
we encounter only a finite number of single poles located at
$u_0=1,\,\,2$ and $3$.

In the same manner we get from Eq.\ (\ref{eq:4.11})

$$
Q^2F_{\eta^{\prime}g^{*}g^{*}}^q(Q^2,\omega=0)=
-\frac{64\pi^2C(1+A(Q^2))}{\beta_0}
\int_0^{\infty}due^{-ut}R(u,t)\frac{1}{(2-u)(3-u)}
$$
\begin{equation}
\label{eq:4.13}
+\frac{640\pi^2CA(Q^2)}{\beta_0}
\int_0^{\infty}due^{-ut}R(u,t)\frac{1}{(3-u)(4-u)(5-u)}.
\end{equation}
In this case the single IR renormalon poles are located at the points
$u_0=2,\;3,\;4$ and $5$.

 In order to get the IR renormalon structure of the integrands in
Eq.\ (\ref{eq:4.4}), we have to expand the hypergeometric function
${}_2F_1(a,b;c;z)$ in powers of $z$ (see Eq.\ (\ref{eq:4.3})), where
the condition $\left| z\right|<1$ must hold.
But the argument of the function $_2F_1(a,b;c;2\omega /(1+\omega ))$
satisfies this requirement only in the region $\omega \in (0,1)$.
In the region $\omega \in (-1,0)$ an expression obtained from
(\ref{eq:4.4})
by means of a simple transformation (see, (\ref{eq:4.18a}) below)
has to be used because in this case the argument of the hypergeometric
function becomes equal to $ 2 \omega/(\omega-1)<1$, obeying in the
region $ \omega \in (-1,0)$ the required constraint.
But regardless of the expansion region, we obtain in both cases the
same IR renormalon structure.
Adding to this argumentation the evident fact that the vertex function
is symmetric under the replacement $\omega \leftrightarrow -\omega$
(cf. Eq.~(\ref{eq:2.31})), we can restrict the study of the form factor
$Q^2F_{\eta^{\prime}g^{*}g^{*}}^q(Q^2,\omega)$ to the region
$\omega \in (0,1)$.
Then, we can expand the hypergeometric functions
$_2F_1(a,b;c;2\omega /(1+\omega ))$ in the region $\omega \in (0,1)$,
via Eq.\ (\ref{eq:4.3}).
For example, for one of these functions, we get
$$
B(2,2-u)_2F_1\left( 1,2;4-u;\beta \right)
=\frac{\Gamma (2)\Gamma (2-u)}{
\Gamma (4-u)}\sum_{k=0}^\infty
\frac{(1)_k(2)_k}{(4-u)_k}\frac{\beta^k}{k!}
$$

$$
=\sum_{k=0}^\infty
\frac{\Gamma (k+2)\Gamma (2-u)}{\Gamma (k+4-u)}\beta^k
=\sum_{k=0}^\infty B(k+2,2-u)\beta ^k
$$
with
$$
\beta= \frac{2\omega}{1+\omega}.
$$
The remaining terms in Eq.\ (\ref{eq:4.4}) can be treated in the same
manner and as a result we obtain
$$
Q^2F_{\eta ^{\prime }g^{*}g^{*}}^q(Q^2,\omega)=-\frac{32\pi
^2C[1+A(Q^2)]}{\beta _0(1+\omega )}\int_0^\infty due^{-ut}R(u,t)
$$

$$
\times \sum_{k=0}^\infty \left[ B(2-u,k+2)+B(2,k+2-u)\right] \beta^k
$$
\begin{equation}
\label{eq:4.14}
+\frac{160\pi ^2CA(Q^2)}{\beta _0(1+\omega )}\int_0^\infty
due^{-ut}R(u,t)\sum_{k=0}^\infty
\left[ B(3-u,k+3)+B(3,k+3-u)\right]\beta^k.
\end{equation}
The IR renormalon structure of the integrands in Eq.\ (\ref{eq:4.14})
is quite clear now.
In fact, we can write the Beta functions entering Eq.\ (\ref{eq:4.14})
in the following form
$$
B(2,k+2-u)=\frac{\Gamma (2)\Gamma (k+2-u)}{\Gamma (k+4-u)}
=\frac 1{(k+2-u)(k+3-u)}
$$
and correspondingly
$$
B(2-u,k+2)=\frac{\Gamma (k+2)}{(2-u)(3-u)\dots (k+3-u)},\,\,
$$

$$
B(3,k+3-u)=\,\frac 2{(k+3-u)(k+4-u)(k+5-u)},
$$

$$
B(3-u,k+3)=\frac{\Gamma (k+3)}{(3-u)(4-u)\dots (k+5-u)}.
$$
Here we have an infinite number of single IR renormalon poles located
at the points
$u_0=k+2,\,k+3;\,\,\,u_0=2,\,3,\ldots k+3;\,\,\,\,u_0=k+3,\,k+4,\,k+5$
and $u_0=3,\,4,\ldots k+5$, respectively.

The last question to be answered is, whether one can use our results,
obtained in the context of the RC method, in the limit
$Q^2\rightarrow \infty$ in order to regain the asymptotic form of the
form factor
$F_{\eta ^{\prime }g^{*}g^{*}}^q(Q^2\rightarrow \infty ,\omega)$.
It is clear that regardless of the methods employed and the
approximations done, in the limit $Q^2\rightarrow \infty $ the form
factor $F_{\eta ^{\prime}g^{*}g^{*}}^q(Q^2,\omega)$ must reach its
asymptotic form.
This is true, of course, for our computations, as we estimate
power-suppressed corrections to the form factor
$Q^2F_{\eta ^{\prime}g^{*}g^{*}}^q(Q^2,\omega )$, which become
important in a region of $Q^2\sim $ of a few GeV${}^2$, but vanish in
the asymptotic limit.
As we have emphasized in Sect.\ II, in the asymptotic limit the
gluon DA of the $\eta ^{\prime}$-meson satisfies
$\phi ^g(x,Q^2)\rightarrow 0$ and hence
$$
Q^2F_{\eta ^{\prime }g^{*}g^{*}}^g(Q^2,\omega)\stackrel{
Q^2\rightarrow \infty }{\longrightarrow }0.
$$
The DA of the quark component $\phi ^q(x,Q^2)$ of the $\eta^{\prime}$
meson evolves for $Q^2\to\infty$ to the asymptotic DA (\ref{eq:2.41})
and all non-asymptotic terms in $\phi ^q(x,Q^2)$ proportional to
$C_n^{3/2}(x-\overline{x}),\,\,n>0$ (in our case $\sim A(Q^2)$)
vanish.
Therefore, the results, which we shall obtain here, describe not only
the asymptotic limit of
$Q^2F_{\eta^{\prime}g^{*}g^{*}}^q(Q^2,\omega)$, but also the asymptotic
limit of the vertex function
$Q^2F_{\eta ^{\prime }g^{*}g^{*}}(Q^2,\omega)$ itself.

In the limit $Q^2 \to \infty$ the asymmetry parameter can take values
$\omega \rightarrow \pm 1$ (if we pass to the limit
$Q^2\rightarrow \infty $ at fixed $Q_2^2$ or $Q_1^2$), $\omega=0$
($Q_1^2=Q_2^2$ and $Q^2 \to \infty$) or $\omega \neq \pm 1,\; 0$ (if
we take the limit $Q^2\rightarrow \infty $ at fixed $\omega $).
We consider here all  possibilities: a)
$Q^2\rightarrow \infty ,\,\,\omega \rightarrow \pm
1$, b) $Q^2 \to \infty, \; \omega=0$ and c)
$Q^2\rightarrow \infty ,\,\,\omega \neq \pm 1,\; 0$.
In the limit $Q^2\rightarrow \infty $, we also take into account that
the second term in the expansion $\alpha_{\rm s}(Q^2 x )$
(\ref{eq:3.8}) has to be neglected.
In other words, in the limit $Q^2\rightarrow \infty $, we find
\begin{equation}
\label{eq:4.15}\int_0^\infty e^{-ut}R(u,t)du\rightarrow \int_0^\infty
e^{-ut}du.
\end{equation}

We begin from the simpler case a). In Eq.\ (\ref{eq:4.12}) we have
already obtained the desired limit, but $A(Q^2)\neq 0$.
Taking into account Eq.\ (\ref{eq:4.15}) and $A(Q^2)\rightarrow 0$,
we get
$$
Q^2F_{\eta ^{\prime }gg^{*}}(Q^2,\omega=\pm 1
)\stackrel{Q^2\rightarrow
\infty}{\longrightarrow}-\frac{16\pi ^2C}{\beta_0}\int_0^\infty \frac{
e^{-ut}du}{1-u}=-\frac{16\pi ^2C}{\beta _0}\frac{li(\lambda )}\lambda .
$$
It is easy to show that using only the leading term in the expansion of
$li(\lambda)/\lambda$ (see, Eq.\ (\ref{eq:3.15})), we find
\begin{equation}
\label{eq:4.16}Q^2F_{\eta ^{\prime }gg^{*}}(Q^2,\omega=\pm 1
)\stackrel{
Q^2\rightarrow \infty }{\longrightarrow }-4\pi C\alpha_{\rm s}(Q^2).
\end{equation}
The limit $Q^2 \to \infty, \; \omega=0$ can be analyzed by similar
means.
Thus, from Eq.\ (\ref{eq:4.13}) we get
$$
Q^2F_{\eta^{\prime}g^{*}g^{*}}(Q^2,\omega=0)
\stackrel{Q^2 \to \infty}{\longrightarrow}
-\frac{64\pi^2C}{\beta_0}\int_0^{\infty}\frac{e^{-ut}du}{(2-u)(3-u)}
$$
$$
=-\frac{64\pi^2C}{\beta_0}\left [\frac{li(\lambda^2)}{\lambda^2}
-\frac{li(\lambda^3)}{\lambda^3} \right ],
$$
which in the limit under consideration simplifies to
\begin{equation}
\label{eq:4.17}
Q^2F_{\eta^{\prime}g^{*}g^{*}}(Q^2,\omega=0)
\stackrel{Q^2\to\infty}{\longrightarrow}
-\frac{8\pi C \alpha_{\rm s}(Q^2)}{3}.
\end{equation}

Now let us consider the more interesting case c).
Then, from Eqs.\ (\ref{eq:4.4})
and (\ref{eq:4.15}), we obtain
$$
Q^2F_{\eta ^{\prime }g^{*}g^{*}}(Q^2,\omega
)\stackrel{Q^2\rightarrow
\infty }{\longrightarrow }-\frac{32\pi ^2C}{\beta _0(1+\omega )}
\int_0^\infty due^{-ut}B(2,2-u)
$$

\begin{equation}
\label{eq:4.18}
\times \left[ \,_2F_1\left( 1,2;4-u;\frac{2\omega }{1+\omega}
\right)+{}_2F_1\left(1,2-u;4-u;\frac{2\omega}{1+\omega}\right)\right] .
\end{equation}
As an example, we analyze the second term in Eq.\ (\ref{eq:4.18})
(see also Eq.\ (\ref{eq:4.14}))
$$
\int_0^\infty due^{-ut}B(2,2-u)_2F_1\left( 1,2-u;4-u;\beta \right)
=\int_0^\infty due^{-ut}\sum_{k=0}^\infty B(2,k+2-u)\beta ^k
$$
$$
=\sum_{k=0}^\infty \Gamma (2)\beta ^k
\int_0^\infty due^{-ut}\left( \frac
1{k+2-u}-\frac 1{k+3-u}\right)
=\sum_{k=0}^\infty \Gamma (2)\left[ \frac{
li(\lambda ^{k+2})}{\lambda ^{k+2}}-\frac{li(\lambda ^{k+3})}{\lambda
^{k+3}}
\right] \beta ^k.
$$
In the considered limit, one finds
$$
\frac{li(\lambda ^{k+2})}{\lambda ^{k+2}}-\frac{li(\lambda
^{k+3})}{\lambda
^{k+3}}\rightarrow \frac 1{\ln \lambda }\left( \frac 1{k+2}-\frac
1{k+3}\right) =\frac 1{\ln \lambda }\frac 1{(k+2)(k+3)}.
$$
Then, the result reads
$$
\sum_{k=0}^\infty \Gamma (2)\left[ \frac{li(\lambda ^{k+2})}{\lambda
^{k+2}}
- \frac{li(\lambda ^{k+3})}{\lambda ^{k+3}}\right] \beta ^k\rightarrow
\frac
1{\ln \lambda }\sum_{k=0}^\infty \frac{\Gamma (2)}{(k+2)(k+3)}\beta^k
$$
$$
=\frac 1{\ln \lambda }\sum_{k=0}^\infty B(2,k+2)\beta ^k=\frac 1{\ln
\lambda
}B(2,2)_2F_1(1,2;4;\beta ).
$$
The same method can be applied to the first function in
Eq.\ (\ref{eq:4.18}).
But before doing that, it is instructive to employ the transformation
\begin{equation}
\label{eq:4.18a}
{}_2F_1(a,b,c;z)
=(1-z)^{-a}{}_2F_1 \left ( a,c-b,c;\frac{z}{z-1}\right ).
\end{equation}
After performing all these operations we find
$$
Q^2F_{\eta ^{\prime }g^{*}g^{*}}(Q^2,\omega
)\stackrel{Q^2\rightarrow
\infty }{\longrightarrow }-\frac{64\pi ^2C}{\beta _0(1+\omega )}\frac
1{\ln
\lambda }B(2,2)_2F_1(1,2;4;\beta )
$$
\begin{equation}
\label{eq:4.19}=-\frac{8\pi C\alpha_{\rm s}(Q^2)}{3(1+\omega
)}\,_2F_1(1,2;4;\beta
).
\end{equation}
Taking into account the expression for the hypergeometric function
$_2F_1(1,2;4;\beta )$ in terms of elementary ones \cite{pr2}

\begin{equation}
\label{eq:4.20}_2F_1(1,2;4;\beta )
=\frac 3{\beta ^3}\left[ \beta (2-\beta
)+2(1-\beta )\ln (1-\beta )\right] ,
\end{equation}
we finally obtain
\begin{equation}
\label{eq:4.21}Q^2F_{\eta ^{\prime }g^{*}g^{*}}(Q^2,\omega)\stackrel{
Q^2\rightarrow \infty }{\longrightarrow }-4\pi C\alpha_{\rm
s}(Q^2)f(\omega
),\,\,\,\,\,\,\,f(\omega )
=\frac 1{\omega ^2}\left( 1+\frac{1-\omega ^2}{
2\omega }\ln \frac{1-\omega }{1+\omega }\right) .
\end{equation}
In our argumentations we have tacitly assumed that $\omega \in (0,1)$.
But Eq.\ (\ref{eq:4.21}) holds for all values of
$\omega \neq \pm 1,\; 0$, which
is evident from the equality $f(\omega)=f(-\omega)$.

Equations (\ref{eq:4.16}), (\ref{eq:4.17}) and (\ref{eq:4.21}) can be
obtained in the standard HSA using the corresponding hard-scattering
amplitudes (\ref{eq:4.9}), (\ref{eq:4.10}) and (\ref{eq:2.21}).
The analysis carried out so far proves the correctness of the RC method
leading to the expected expressions for the form factor
$Q^2F_{\eta^{\prime }g^{*}g^{*}}(Q^2,\omega)$ at
$Q^2\rightarrow \infty $ and demonstrating at the same time the
consistency of the symmetrization of the hard-scattering amplitudes
(\ref{eq:2.33}), (\ref{eq:2.34}).

\subsection{Contribution of the gluon component of the
            $\mathbf{\eta^{\prime}}$-meson to the form factor
            $\mathbf{F_{\eta^{\prime}g^{*}g^{*}}(Q^2,\omega)}$}
The contribution of the gluon component of the $\eta ^{\prime }$-meson
to the form factor $F_{\eta ^{\prime }g^{*}g^{*}}(Q^2,\omega)$ can be
computed using the methods described in the previous subsection.

Using Eq.\ (\ref{eq:2.34}) for the hard-scattering amplitude
$T_1^g(x,Q^2,\omega)$, Eq.\ (\ref{eq:2.28}) and
the gluon component of the $\eta^{\prime}$-meson DA, we have for the
gluon part of the form factor

$$
Q^2F_{\eta^{\prime}g^{*}g^{*}}^g(Q^2,\omega)
=\frac{4\pi^2CB(Q^2)}{3\beta_0}
\int_{0}^{\infty}due^{-ut}R(u,t)
\left [\int_{0}^{1}dxx^{1-u}\overline{x}(x-\overline{x})\right .
$$

\begin{equation}
\label{eq:4.22}
\left . \frac{(1+\omega)x+
(1-\omega)\overline{x}}{\omega
\left [(1+\omega)\overline{x}+(1-\omega)x \right ]}
+\int_{0}^{1}dxx\overline{x}^{1-u}(x-\overline{x})
\frac{(1+\omega)x+(1-\omega)\overline{x}}{\omega \left [
(1+\omega)\overline{x}+(1-\omega)x \right ]}\right ] ,
\end{equation}
which after some simple calculations becomes
$$
Q^2F_{\eta ^{\prime }g^{*}g^{*}}^g(Q^2,\omega)=\frac{4\pi
^2CB(Q^2)
}{3\beta _0\omega}\int_0^\infty due^{-ut}R(u,t)\left\{
B(4-u,2){}_2F_1 \left (1,4-u;6-u;\frac{2\omega}{1+\omega}\right )\right.
$$

$$
+B(4,2-u)\,_2F_1\left (1,4;6-u;\frac{2\omega}{1+\omega} \right )
$$

$$
-\frac{2\omega }{1+\omega }B(3,3-u)\left[
_2F_1 \left (1,3-u;6-u;\frac{2\omega}{1+\omega}
\right )+{}_2F_1 \left (1,3;6-u;\frac{2\omega}{1+\omega} \right ) \right]
$$

\begin{equation}
\label{eq:4.23}\left . -\frac{1-\omega }{1+\omega } \left[
B(2-u,4)_2F_1 \left (1,2-u;6-u;\frac{2\omega}{1+\omega}
\right )+B(2,4-u)_2F_1 \left (1,2;6-u;\frac{2\omega}{1+\omega}\right )\right]
\right\} .
\end{equation}

For the $\eta ^{\prime }$-meson on-shell-gluon transition, we find
$$
Q^2F_{\eta ^{\prime }gg^{*}}^g(Q^2,\omega =\pm 1)=\frac{4\pi
^2CB(Q^2)}{3
\beta _0}\int_0^\infty due^{-ut}R(u,t)
$$

\begin{equation}
\label{eq:4.25}
\times \left[ B(1,4-u)+B(4,1-u)-B(2,3-u)-B(3,2-u)\right] ,
\end{equation}
or equivalently
\begin{equation}
\label{eq:4.26}Q^2F_{\eta ^{\prime }gg^{*}}^g(Q^2,\omega =\pm 1)=
\frac{4\pi
^2CB(Q^2)}{3\beta _0}
\int_0^\infty due^{-ut}R(u,t)\left( \frac 1{1-u}-\frac
4{2-u}+\frac 4{3-u}\right) .
\end{equation}

The form factor
$Q^2F_{\eta^{\prime}g^{*}g^{*}}^g(Q^2, \omega=0)$
can be calculated by employing the following form for the
hard-scattering amplitude
\begin{equation}
\label{eq:4.27}
T_1^g(x, Q^2, \omega)-T_2^g(x,Q^2,\omega)=\frac{2\pi}{3Q^2}
\left [\alpha_{\rm s}(Q^2x)+\alpha_{\rm s}(Q^2\overline{x})\right ]
\frac{x-\overline{x}}{1-\omega^2(x-\overline{x})^2},
\end{equation}
which for $\omega=0$ leads to a very simple expression.
Then, it is not difficult to demonstrate that
$$
Q^2F_{\eta^{\prime}g^{*}g^{*}}^g(Q^2,\omega=0)
=\frac{16\pi^2CB(Q^2)}{3\beta_0}
\int_0^{\infty}due^{-ut}R(u,t)\left [B(4-u,2) \right .
$$
\begin{equation}
\label{eq:4.28}
\left .-2B(3-u,3)+B(2-u,4)\right ].
\end{equation}
This latter expression can be recast into the form
$$
Q^2F_{\eta^{\prime}g^{*}g^{*}}^g(Q^2,\omega=0)
=\frac{16\pi^2CB(Q^2)}{3\beta_0}
\int_0^{\infty}due^{-ut}R(u,t)\left [\frac{1}{2-u}\right .
$$
\begin{equation}
\label{eq:4.29}
\left .-\frac{5}{3-u}+\frac{8}{4-u}-\frac{4}{5-u}\right ].
\end{equation}

The IR renormalon structure of the integrands in Eq.\ (\ref{eq:4.26})
and (\ref{eq:4.29}) is obvious:
they have a finite number of IR renormalon poles located at the points
$u_0=1,2,3$ and $u_0=2, 3, 4, 5$, respectively.
In order to find the IR renormalon structure of the integrand in
Eq.\ (\ref{eq:4.23}), we expand the corresponding hypergeometric
functions over $\beta =2\omega/(1+\omega )$ in the region
$\omega \in (0,1)$, providing the following result

$$
Q^2F_{\eta ^{\prime }g^{*}g^{*}}^g(Q^2,\omega)
=
\frac{4\pi^2CB(Q^2)}{3\beta _0\omega}
\int_0^\infty due^{-ut}R(u,t)\sum_{k=0}^\infty \left\{ \left[
B(k+4,2-u)\right. \right.
$$

$$
\left. +B(k+4-u,2)\right] \beta ^k-\frac{1-\omega }{1+\omega }\left[
B(k+2,4-u)+B(k+2-u,4)\right] \beta ^k
$$
\begin{equation}
\label{eq:4.30}\left. -\left[ B(k+3,3-u)+B(k+3-u,3)\right] \beta
^{k+1}\right\}.
\end{equation}
Using the identities

$$
B(k+4,2-u)=\frac{\Gamma(k+4)}{(2-u)(3-u)\ldots(k+5-u)},
$$

$$
B(k+4-u,2)=\frac{1}{(k+4-u)(k+5-u)},
$$

$$
B(k+2,4-u)=\frac{\Gamma(k+2)}{(4-u)(5-u)\ldots(k+5-u)},
$$

$$
B(k+2-u,4)=\frac{6}{(k+2-u)(k+3-u)(k+4-u)(k+5-u)},
$$

$$
B(k+3,3-u)=\frac{\Gamma(k+3)}{(3-u)(4-u)\ldots(k+5-u)},
$$

$$
B(k+3-u,3)=\frac{2}{(k+3-u)(k+4-u)(k+5-u)}
$$
it is easy to conclude that there is an infinite number of IR
renormalon poles, being located at the points $u_0=2,3,\ldots k+5$;
$u_0=k+4,\:k+5$; $u_0=4,5,\ldots k+5$; $u_0=k+2,\;k+3,\;k+4,\;k+5$;
$u_0=3,4,\ldots k+5$ and $u_0=k+3,\;k+4,\;k+5$.

\begin{figure}[t]
\centering\epsfig{file=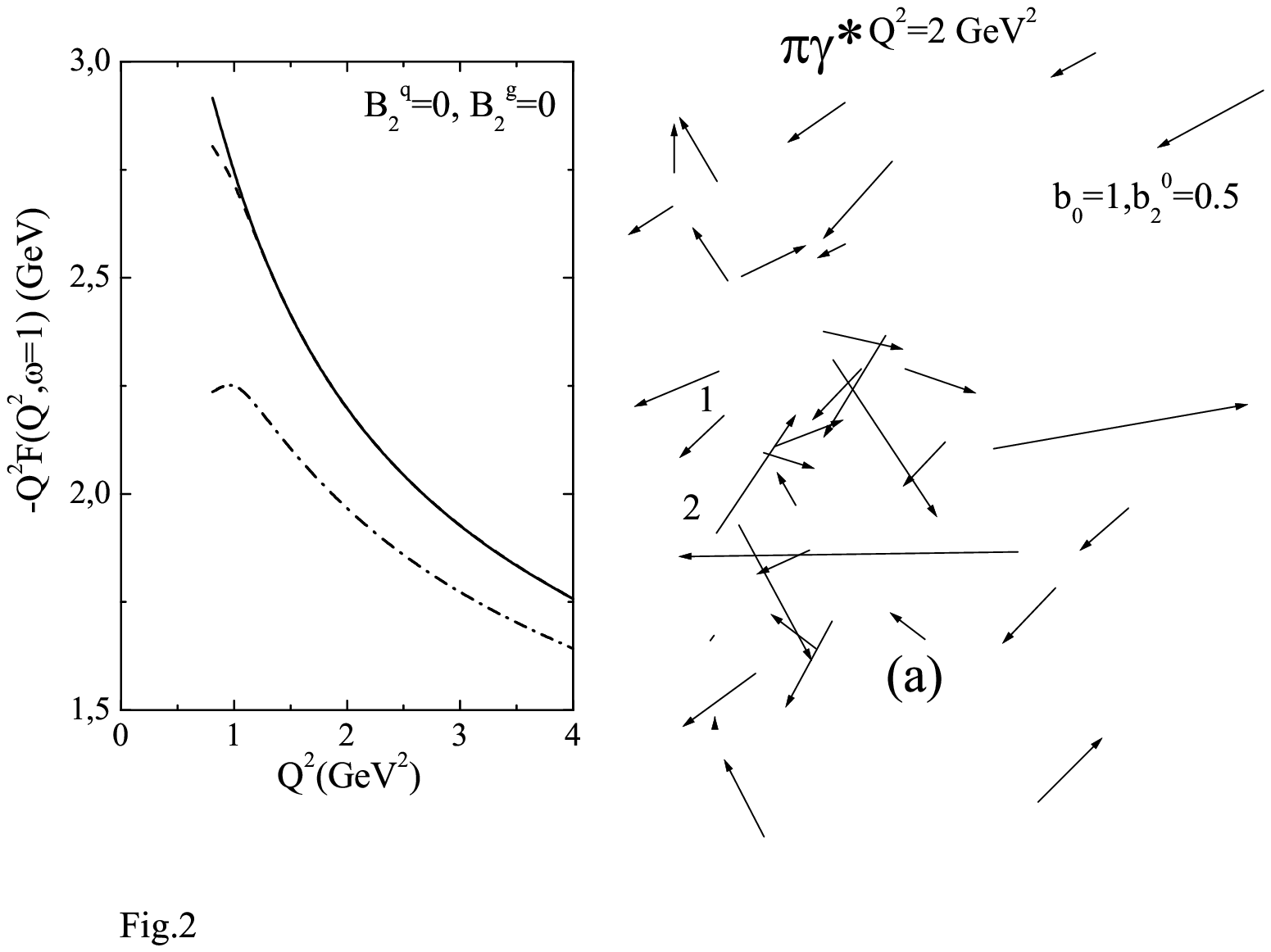,height=8cm,width=6.0cm,clip=}
\caption{
Scaled $\eta^{\prime}g$ transition form
factor
$-Q^2F_{\eta^{\prime}gg^{*}}(Q^2,\omega= \pm 1)$ vs. $Q^2$.
The solid line is computed within the RC method, the dashed one is
found using the IR matching scheme with $f_2(2~{\rm GeV})\simeq 0.535$.
The dot-dashed curve is obtained in the framework of the IR matching
scheme using for $f_2(2~{\rm GeV})$ the experimental value $0.5$.}
\label{fig:gfig3}
\end{figure}

\section{Numerical analysis}
\setcounter{equation}0
We begin this section by comparing the results obtained with the RC
method with those from the IR matching scheme.
In Sect.\ III we have noted that from experimental data only values of
the nonperturbative parameters
$f_1(2~{\rm GeV})$ and $f_2(2~{\rm GeV})$
have been extracted.
We also know that the lowest-order moment integral and hence the
parameter entering our formulas is $f_2(2~{\rm GeV})$.
Therefore, to make the comparison as clear as possible, we should choose
the input parameters for the $\eta^{\prime}$-virtual gluon transition
in such a way as to determine the behavior of the form factor solely
with $f_2(2~{\rm GeV})$.
This can be easily achieved if we set for the $\eta^{\prime}$-meson
DA parameters
$$
B_2^q(\mu_0^2)=0,\:\:\: B_2^g(\mu_0^2)=0.
$$
Under these circumstances, the gluon component of the vertex function
vanishes.
To remove from the analysis the higher-moment integrals
$f_p(Q),\:p >2$, we consider only the $\eta^{\prime}$-meson
on-shell-gluon transition, i.e., the $\omega= \pm 1$ case.
Moreover, we neglect the $\sim \alpha_S^2$ order term in
Eq.\ (\ref{eq:3.2}) and set in Eq.\ (\ref{eq:3.8}) $R(u,t)=1$ because
in Eq.\ (\ref{eq:3.19}) $\alpha_s$ is used at the level of the one-loop
order accuracy.
After these simplifications, the FF is given by the following
expression
\begin{equation}
\label{eq:5.1}
Q^2F_{\eta^{\prime}gg^{*}}(Q^2,\omega= \pm 1)=
-\frac{16 \pi^{2} C}{\beta_0}\int^{\infty}_0
\frac {e^{-ut}du}{1-u}=-4\pi Cf_2(Q).
\end{equation}

\begin{figure}[t]
\centering\epsfig{file=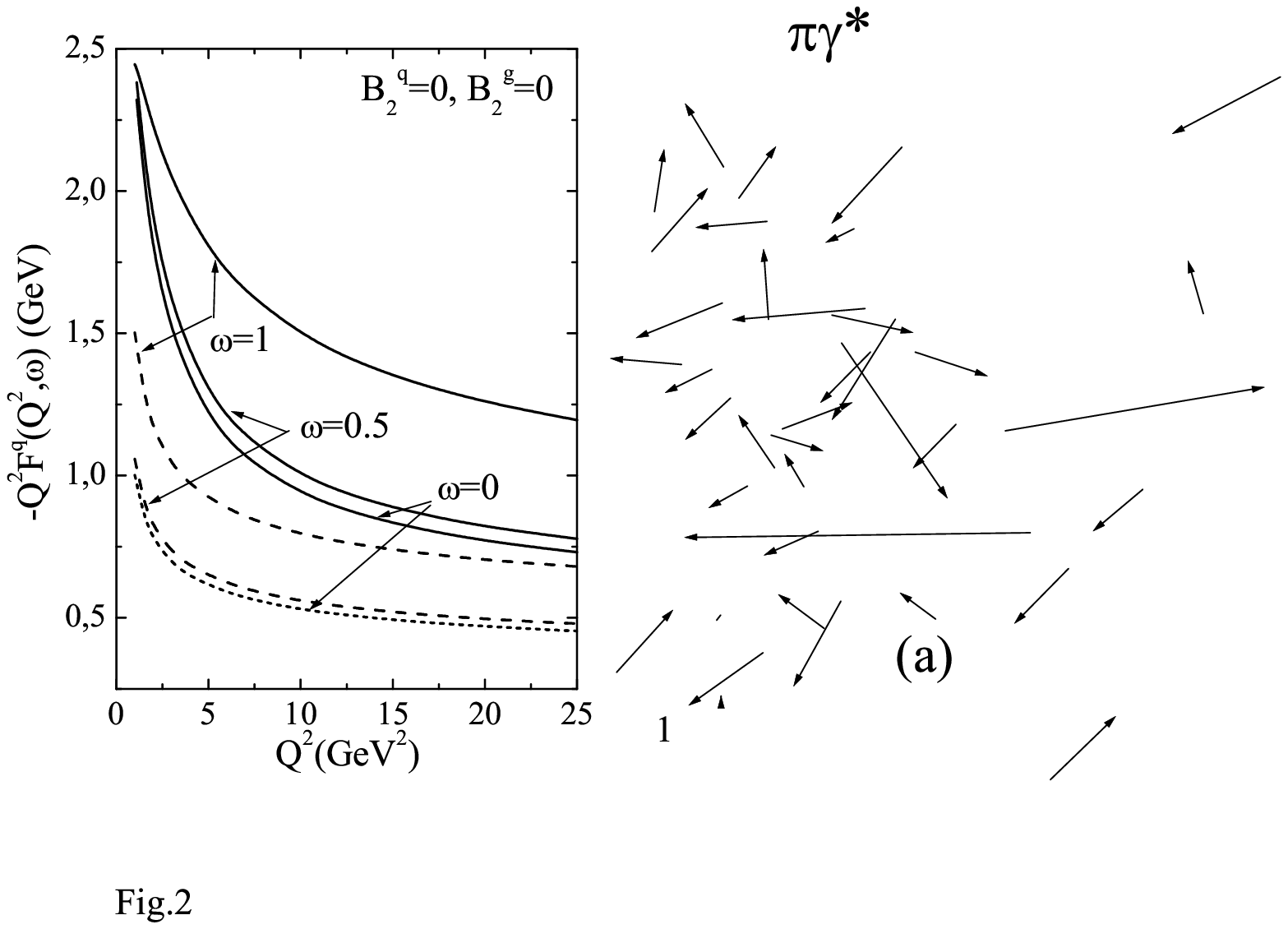,height=8cm,width=6.0cm,clip=}
\hspace{1.5cm}
\centering\epsfig{file=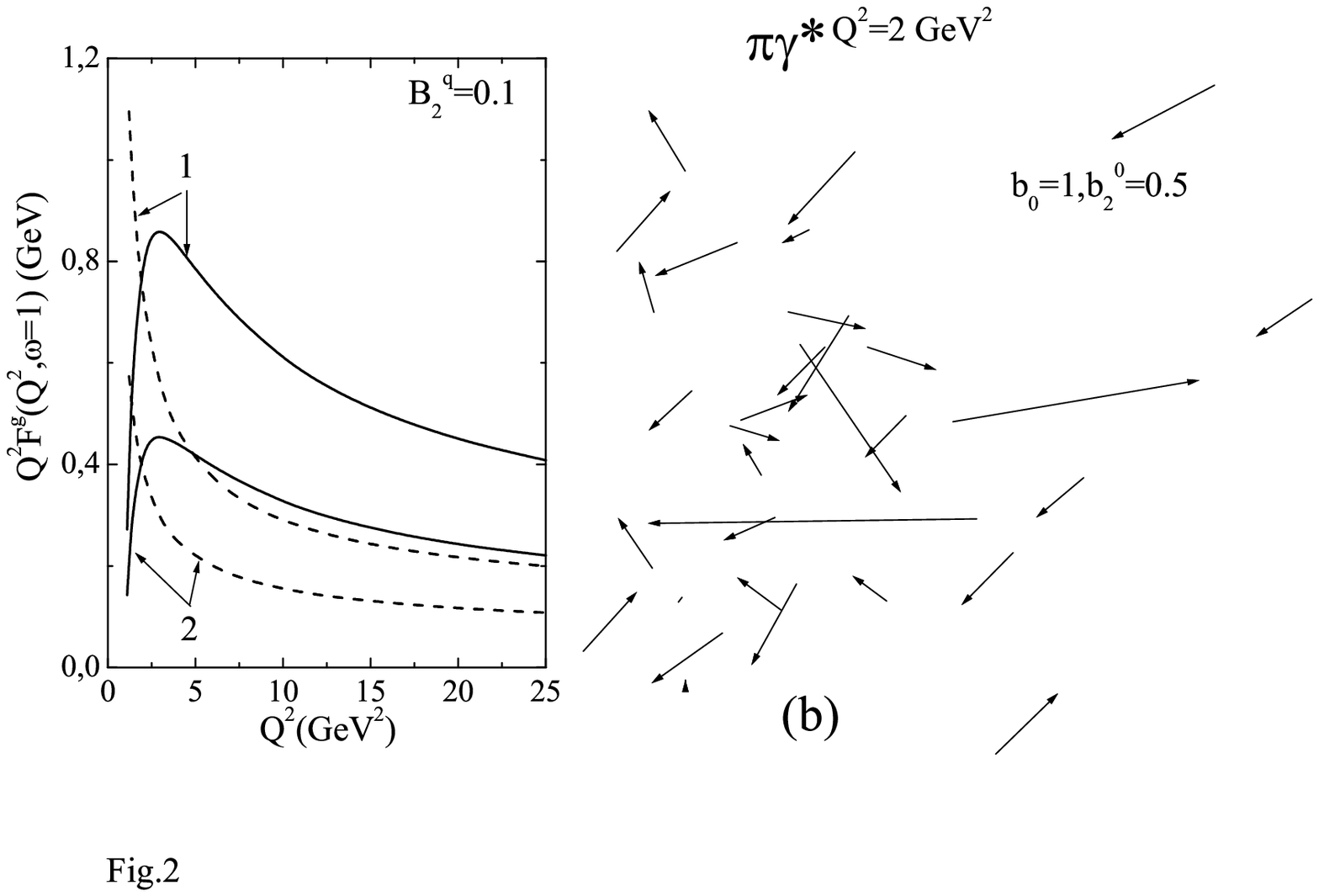,height=8cm,width=6.0cm,clip=}
\vspace{0.0cm}
\caption{
The quark (left panel) and gluon (right panel) components of
the transition form factor
$Q^2F_{\eta ^{\prime}g^{*}g^{*}}^q(Q^2,\omega)$ as functions of $Q^2$.
The solid curves are obtained using the RC method, whereas the
broken lines are calculated within the standard HSA. The quark
component is computed at various $\omega$ values by employing the
asymptotic DA. In the right panel the correspondence between the curves
and the input parameter $B_2^{g}$ is $B_2^{g}=8$ for the curves
labelled $1$ and $B_2^{g}=4$ for those labelled $2$.}
\label{fig:gfig4}
\end{figure}

Results of our computations are shown in Fig.\ \ref{fig:gfig3}, where
in order to distinguish the various curves, these are displayed in the
region of $Q^2 \in [1,4]$~GeV${}^{2}$, whereas at higher $Q^2$ they
are close to each other.
The RC method and the IR matching scheme\footnote{Note that in the
calculations within the IR matching scheme, Eq.\ (\ref{eq:3.19})
with $N=4$ has been used.} both lead almost to identical predictions in
the entire domain $1~{\rm GeV}^2 \leq Q^2 \leq 25~{\rm GeV}^2$,
provided that in the IR matching scheme one uses in expression
(\ref{eq:3.19}) the value $f_2(2~{\rm GeV}) \simeq 0.535$ found within
the RC method (\ref{eq:3.23}).
The curve following from the IR matching scheme deviates from the
prediction obtained with the RC method only for $Q^2<1.4~{\rm GeV}^2$.
On the other hand, the deviation of that curve, calculated using the
experimental value $f_2(2~{\rm GeV})\simeq 0.5$, from the result of
the RC method is sizeable in the region $Q^2 =1\div 2~{\rm GeV}^2$,
reaching $\sim 30 \%$ at $Q^2=1~{\rm GeV}^2$.
A similar behavior was observed in the calculation of the pion
electromagnetic form factor, carried out within the context of these
methods \cite{ag4}.
The difference between the solid and the dot-dashed lines in
Fig.\ \ref{fig:gfig3} is considerably reduced when varying the QCD
scale parameter $\Lambda=0.3~{\rm GeV}$ or the experimental value of
$f_2(2~{\rm GeV})$ within their corresponding uncertainty limits.
But we are not going to make decisive conclusions from these rather
model-dependent calculations.
Our aim here is to check and demonstrate that the RC method and the
IR matching scheme predict in the considered region
$1 ~{\rm GeV}^2 \leq Q^2\leq 25~{\rm GeV}^2$
almost identical results that do not contradict experiment.

\begin{figure}[t]
\centering\epsfig{file=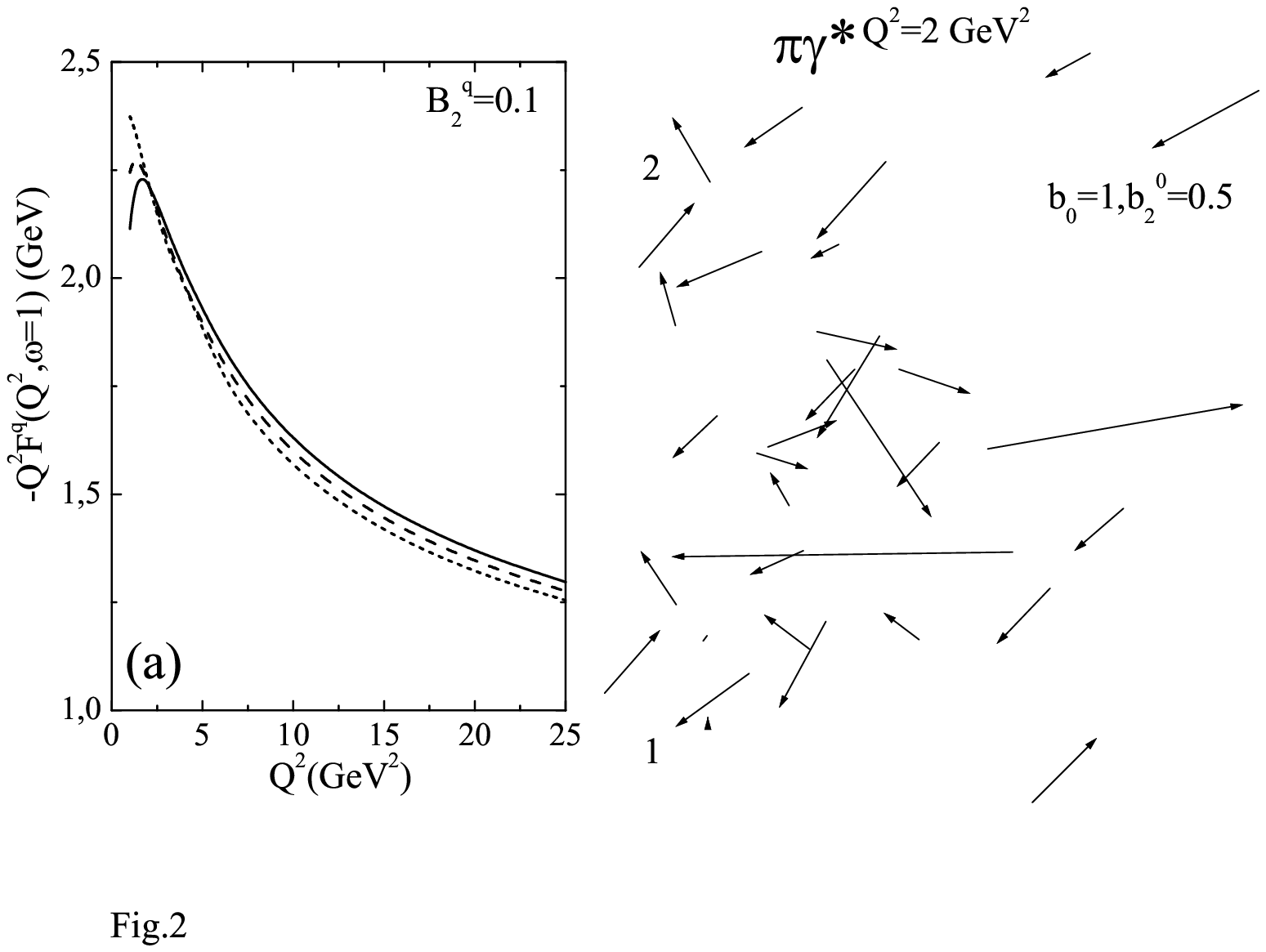,height=8cm,width=6.0cm,clip=}
\hspace{1.5cm}
\centering\epsfig{file=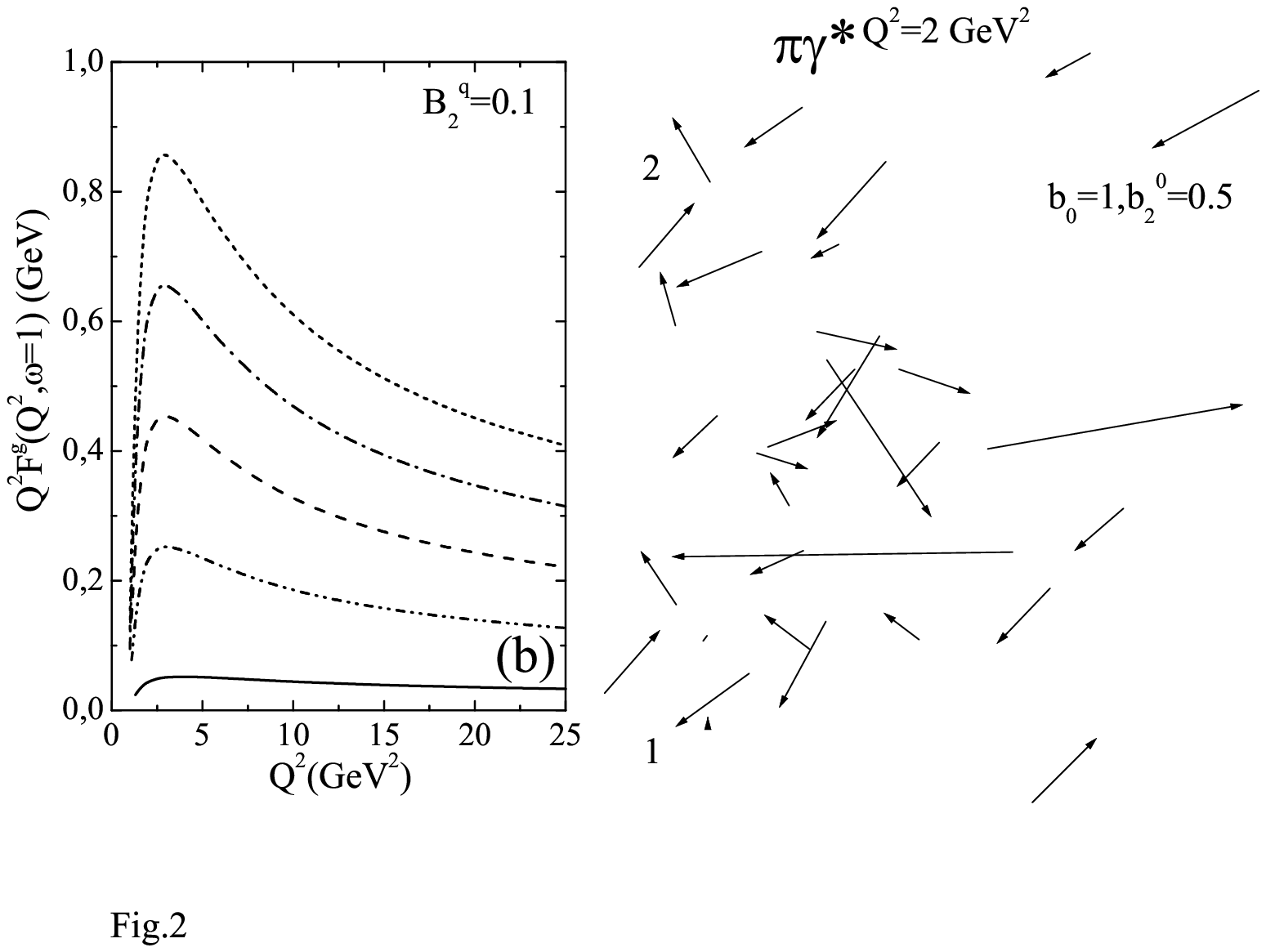,height=8cm,width=6.0cm,clip=}
\vspace{0.0cm}
\caption {The contribution of the quark a) and gluon b)
components of the $\eta ^{\prime}$ meson to the form factor
$Q^2F_{\eta ^{\prime}gg^{*}}(Q^2,\omega=\pm 1)$ vs. $Q^2.$ All curves
are obtained in the context of the RC method. The correspondence
between all displayed curves and the
parameter $B_2^g$ is: $B_2^g=0$ for solid curves; $B_2^g=2$ for
the dot-dot-dashed curve; $B_2^g=4$ for the dashed lines; $B_2^g=6$
for the dot-dashed line, and $B_2^g=8$ for the short-dashed curves.
In figure a) curves corresponding only to $B_2^g=0,\:4$, and $8$
are shown.}
\label{fig:gfig5}
\end{figure}

 In order to proceed with the computation of the $\eta^{\prime}$-meson
gluon vertex function and explore the role played by the
$\eta^{\prime}$ gluon content in this process, we have to define the
allowed values of the free input parameters $B_2^q$ and $B_2^g$ at the
normalization point $\mu_0^2=1\; {\rm GeV}^2$.
These parameters determine the shape of the DA's of the quark and gluon
components of the $\eta^{\prime}$-meson and, in general, they have to
be extracted from experimental data or computed by non-perturbative
techniques.
The comparison of the $\eta^{\prime}$-meson photon electromagnetic
transition FF $F_{\eta^{\prime}\gamma}(Q^2)$ with the CLEO data
leads to the conclusion
that the $\eta^{\prime}$-meson DA must be close to its asymptotic form
with a coefficient $B_2^q \simeq 0.1$ deduced in \cite{ag1}.
But this conclusion was made by neglecting the contribution of the gluon
component of the $\eta^{\prime}$-meson to the $\eta^{\prime}\gamma$
transition FF.
The investigation of the FF $F_{\eta^{\prime}\gamma}(Q^2)$ was extended
and revised in \cite{ag10}. In fact, in this work the FF
$F_{\eta^{\prime}\gamma}(Q^2)$
was computed within the RC method by taking into account contributions
arising due to both the quark and gluon components of the $\eta^{\prime}$
meson DA. The comparison with the CLEO data demonstrated that allowed values
of the Gegenbauer coefficients $B_2^q(1 \; {\rm GeV}^2)$ and
$B_2^g(1 \; {\rm GeV}^2)$ are strongly correlated. They were extracted
in Ref.\ \cite{ag10} and read
$$
B_2^q(1 \; {\rm GeV}^2)=0, \; \; B_2^g(1 \; {\rm GeV}^2) \in [4,18],
$$
\begin{equation}
\label{eq:5.1a}
B_2^q(1 \; {\rm GeV}^2)=0.05, \; \; B_2^g(1 \; {\rm GeV}^2) \in [0,16],
\end{equation}
and
\begin{equation}
\label{eq:5.1b}
B_2^q(1 \; {\rm GeV}^2)=0.1, \; \; B_2^g(1 \; {\rm GeV}^2) \in [-2,14].
\end{equation}
In the present paper we select values of the parameters $B_2^q(1 \; {\rm GeV}^2)$ and
$B_2^g(1 \; {\rm GeV}^2)$ that obey the constraints (\ref{eq:5.1a}) and
(\ref{eq:5.1b}).

It is evident that the non-asymptotic terms in the quark and gluon DA's
of the $\eta^{\prime}$-meson proportional to $A(Q^2)$ and $B(Q^2)$,
respectively, affect the asymptotic value of the
$\eta^{\prime}$-meson-gluon transition form factor.
Therefore, before presenting contributions from these terms, it is
instructive to study the asymptotic FF itself.
In the left panel of Fig.\ \ref{fig:gfig4}, we depict the
$\eta^{\prime}$-meson - virtual gluon transition FF as a function of
the gluon virtuality $Q^2$.
For the asymptotic DA the quark component of the form factor
coincides with the full one.
In the same figure the predictions obtained within the standard
HSA are also shown.
One sees that in the domain
$1\: \rm {GeV}^2 \leq Q^2 \leq 25\: \rm {GeV}^2$ the standard
pQCD results get enhanced by approximately a factor of two due to
power corrections.
A similar conclusion is valid also for the gluon component of
the form factor (right panel in Fig.\ \ref{fig:gfig4}, computed by
employing the $\eta^{\prime}$-meson DA's).

\begin{figure}[t]
\centering\epsfig{file=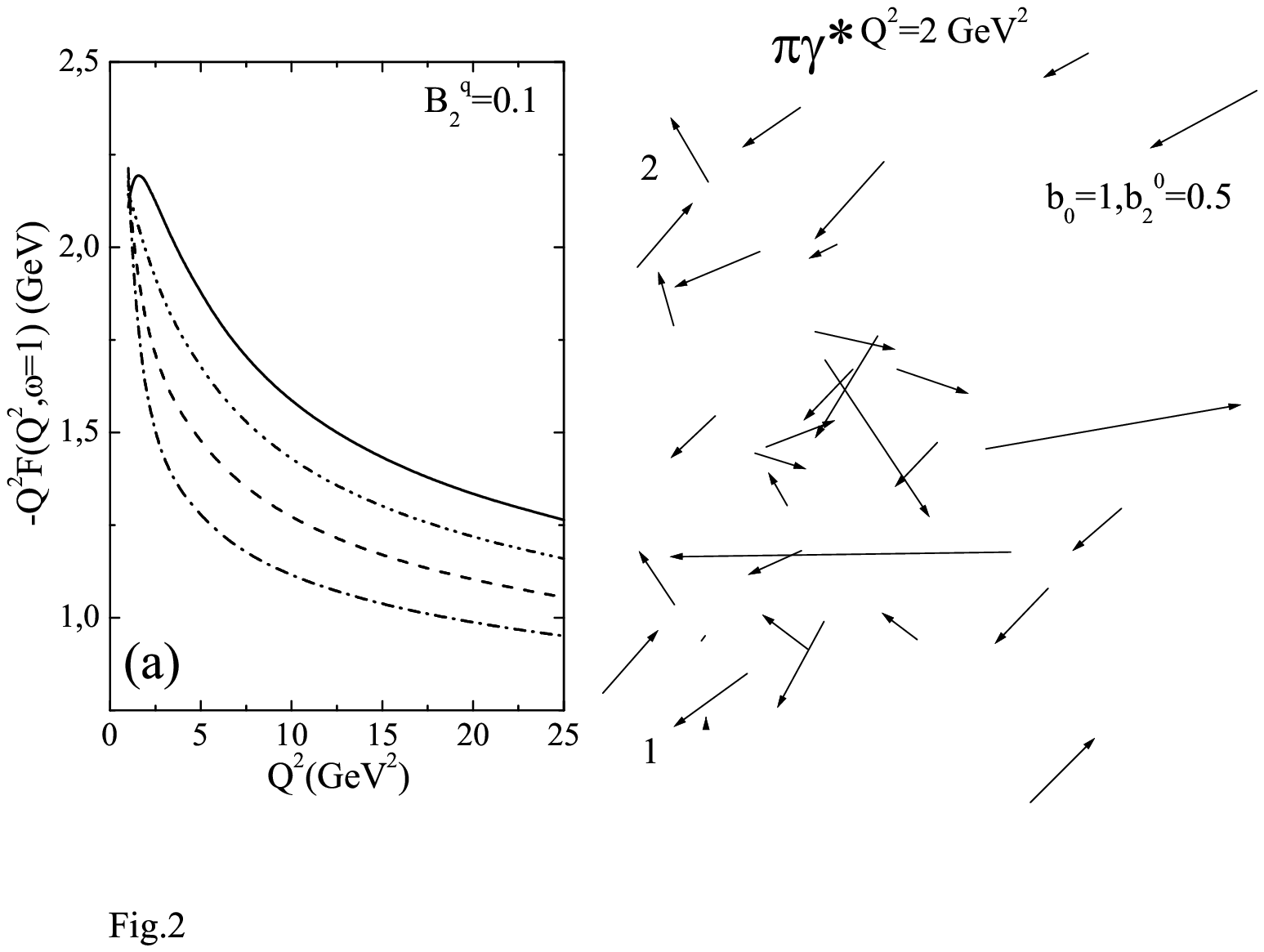,height=8cm,width=6.0cm,clip=}
\hspace{1.5cm}
\centering\epsfig{file=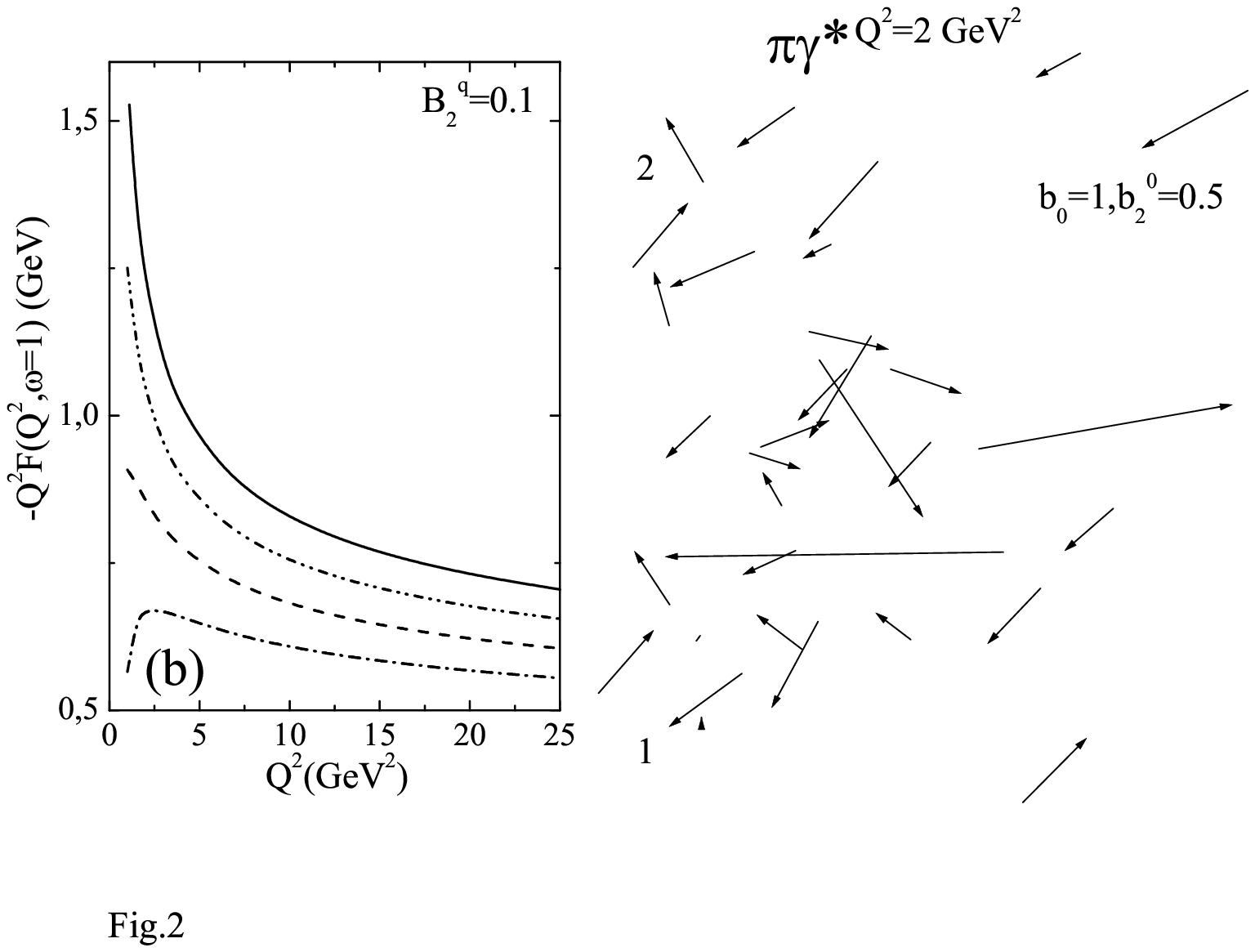,height=8cm,width=6.0cm,clip=}
\vspace{0.0cm}
\caption{The form factor
$-Q^2F_{\eta^{\prime}gg^{*}}(Q^2,\omega=\pm 1)$
computed using the RC method a) and the standard HSA b).
The correspondence
between displayed curves and the
parameter $B_2^g$ is: $B_2^g=0$ for solid curves; $B_2^g=2$ for
the dot-dot-dashed curves; $B_2^g=4$ for the dashed lines; $B_2^g=6$
for the dot-dashed lines.}
\label{fig:gfig6}
\end{figure}

Here some comments concerning the accuracy of the numerical
computations are in order.
In Fig.\ \ref{fig:gfig4} (left panel) and in the following ones, the
curves obtained within the RC method, as a rule, require a summation of
an infinite series.
In real numerical computations we truncate such a series at some
$k=K_{max}$.
Naturally, the question arises about the convergence rate of this
series.
Let us explain this problem by considering Fig.\ \ref{fig:gfig4} as an
example.
The solid curve with $\omega=0.5$ in Fig.\ \ref{fig:gfig4} (left panel)
has been found within the RC method by employing Eq.\ (\ref{eq:4.14}).
This expression contains a series with factorially growing coefficients.
For definiteness we analyze the term
$$
\int_{0}^{\infty}due^{-ut}R(u,t)\sum_{k=0}^{\infty}B(k+2,2-u)\beta^{k}
$$

\begin{equation}
\label{eq:5.2}
 =\sum^{\infty}_{k=0}
\int_{0}^{\infty}due^{-ut}R(u,t)
\frac {\Gamma(k+2)}{(2-u)(3-u)\ldots(k+3-u)}
\beta^{k}.
\end{equation}
The expansion parameter $\beta=2\omega/(1+\omega)$ in
Eq.\ (\ref{eq:5.2}) at the point $\omega=0.5$ is equal to
$\beta=2/3$.
Below, we write down the values of the Gamma function
$\Gamma(k+2)=(k+1)!$ and also those of the product of $\beta^{k}$
with the principal value of the integral
 \begin{equation}
\label{eq:5.3}
I(k)=\int_{0}^{\infty}due^{-ut}R(u,t)
\frac {\beta^{k}}{(2-u)(3-u)\ldots(k+3-u)}
\end{equation}
for
$k=0$, $5$ and $10$

$$
\Gamma(2)=1,\:\: \Gamma(7)=720, \:\: \Gamma(12)=3.99168\cdot10^{7},
$$
and

$$
I(0)\simeq 0.09041, \;\;\; I(5) \simeq 2.44061 \cdot 10^{-6},\;\;
I(10) \simeq 1.31568 \cdot 10^{-12} \; .
$$
As a result, the corresponding terms in the sum given by
Eq.\ (\ref{eq:5.2}) take the values

$$
0.09041, \quad 1.75724 \cdot 10^{-3},\quad  5.25177 \cdot
10^{-5},
$$
respectively. The calculations above have been performed at $Q^2=2\ {\rm
GeV}^2$. At the momentum transfer $Q^2=20\ {\rm GeV}^2$ we get
$$
0.03898, \;\;\; 7.59618 \cdot 10^{-4}, \;\; 4.42997 \cdot 10^{-5}.
$$
One observes that the convergence rates of the numerical series are
high and that we can therefore truncate them, as a rule, at
$K_{\rm max}=20$.

\begin{figure}[t]
\centering\epsfig{file=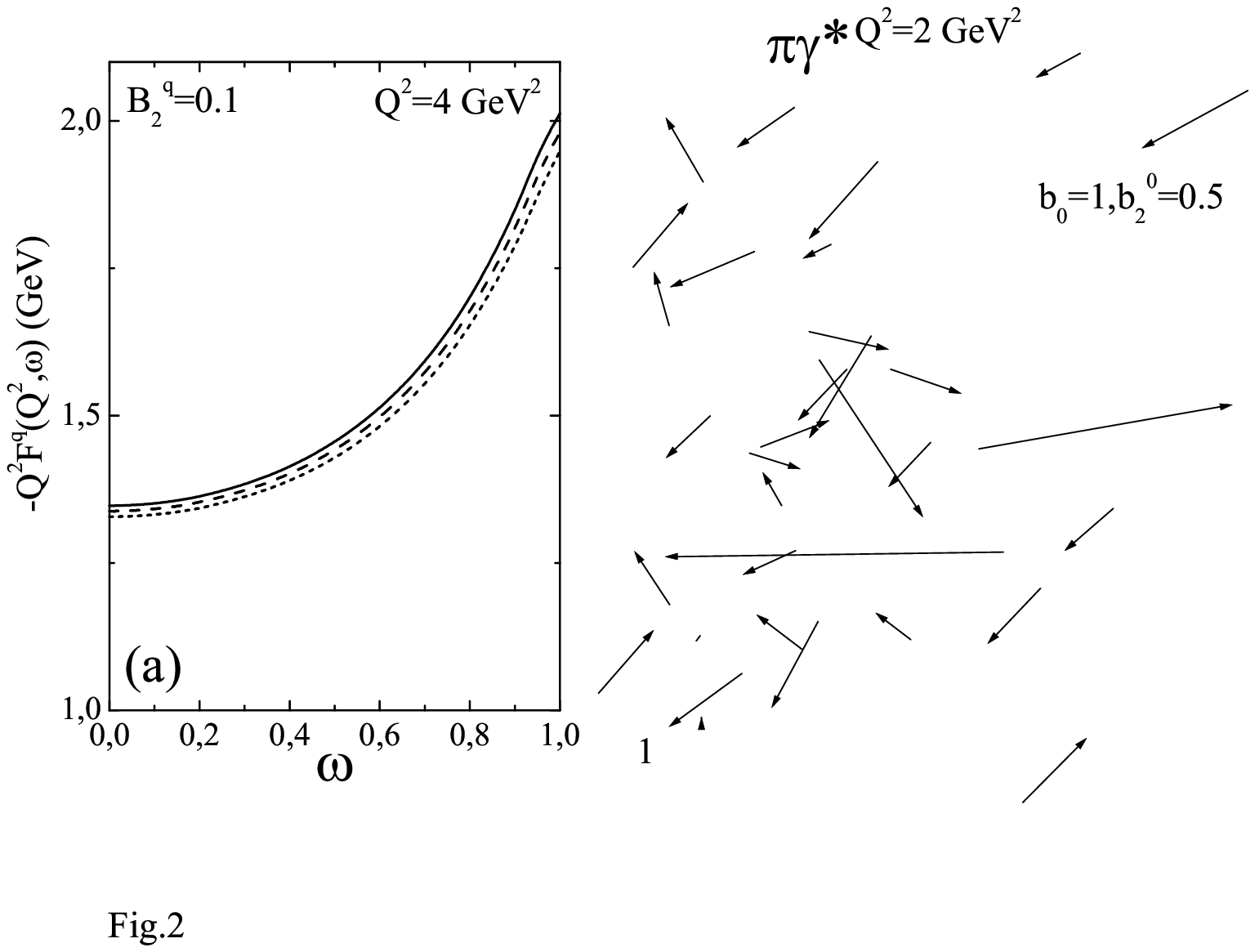,height=8cm,width=6.0cm,clip=}
\hspace{1.5cm}
\centering\epsfig{file=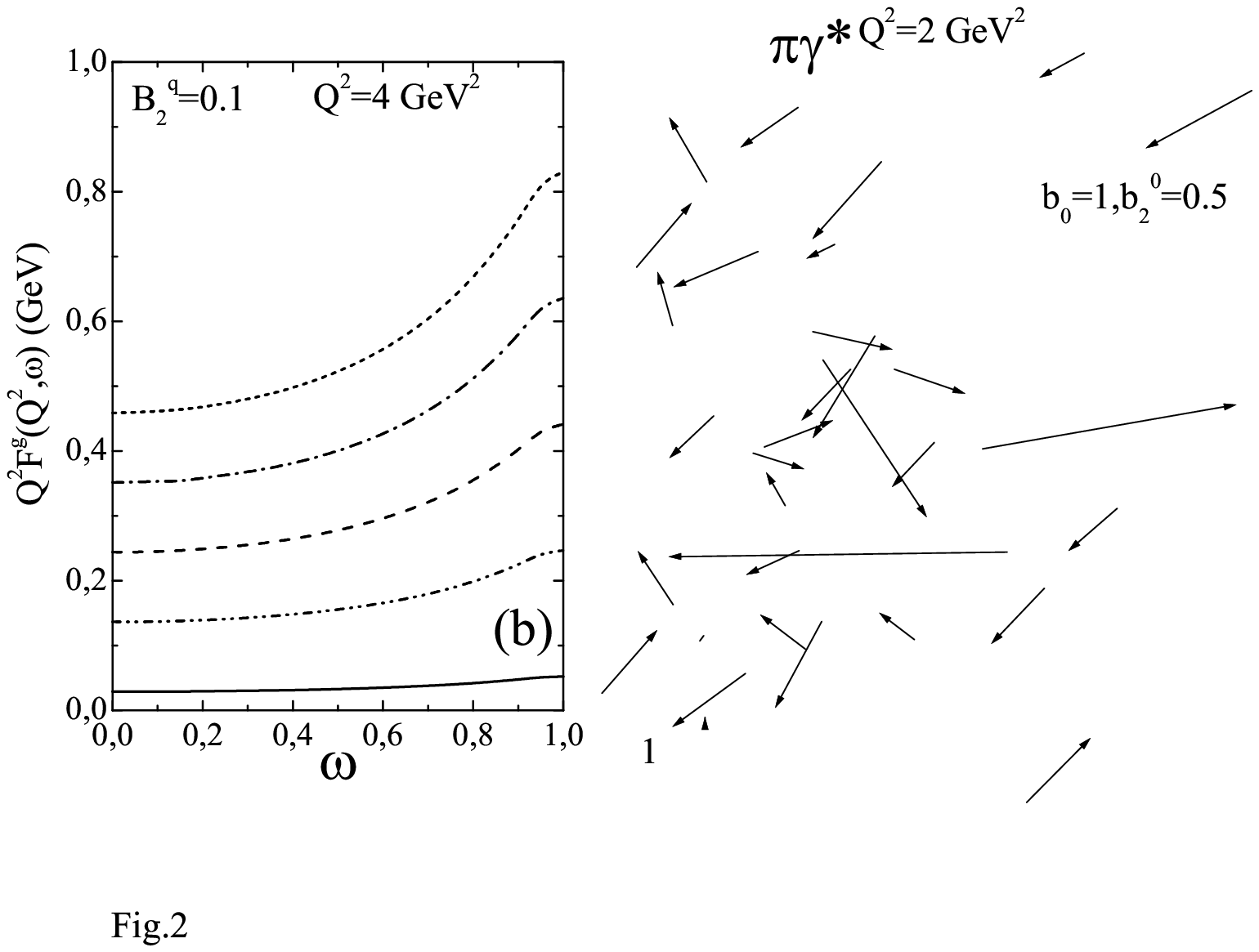,height=8cm,width=6.0cm,clip=}
\vspace{0.0cm}
\caption{The quark a) and gluon b) components of the FF at fixed
total gluon virtuality $Q^2=4\;\rm {GeV}^2$ as functions
of the asymmetry parameter $\omega$. The RC method is employed.
The correspondence between depicted lines and the parameter
$B_2^g$ is the same as in Fig.\ \ref{fig:gfig5}.}
\label{fig:gfig7}
\end{figure}

We have analyzed the impact of the various DA's of the
$\eta^{\prime}$-meson on the
$\eta^{\prime}g$ transition form factor.
The quark component of the FF is stable for different values of
$B_2^g\in [0,8]$.
It is difficult to distinguish the corresponding curves and therefore
in Fig.\ 5a we can only plot some of them.
In contrast, the gluon component of the form factor demonstrates a
rapid growth with $B_2^g$ (Fig.\ 5b).
As a result, due to different signs of the quark
\begin{figure}[t]
\centering\epsfig{file=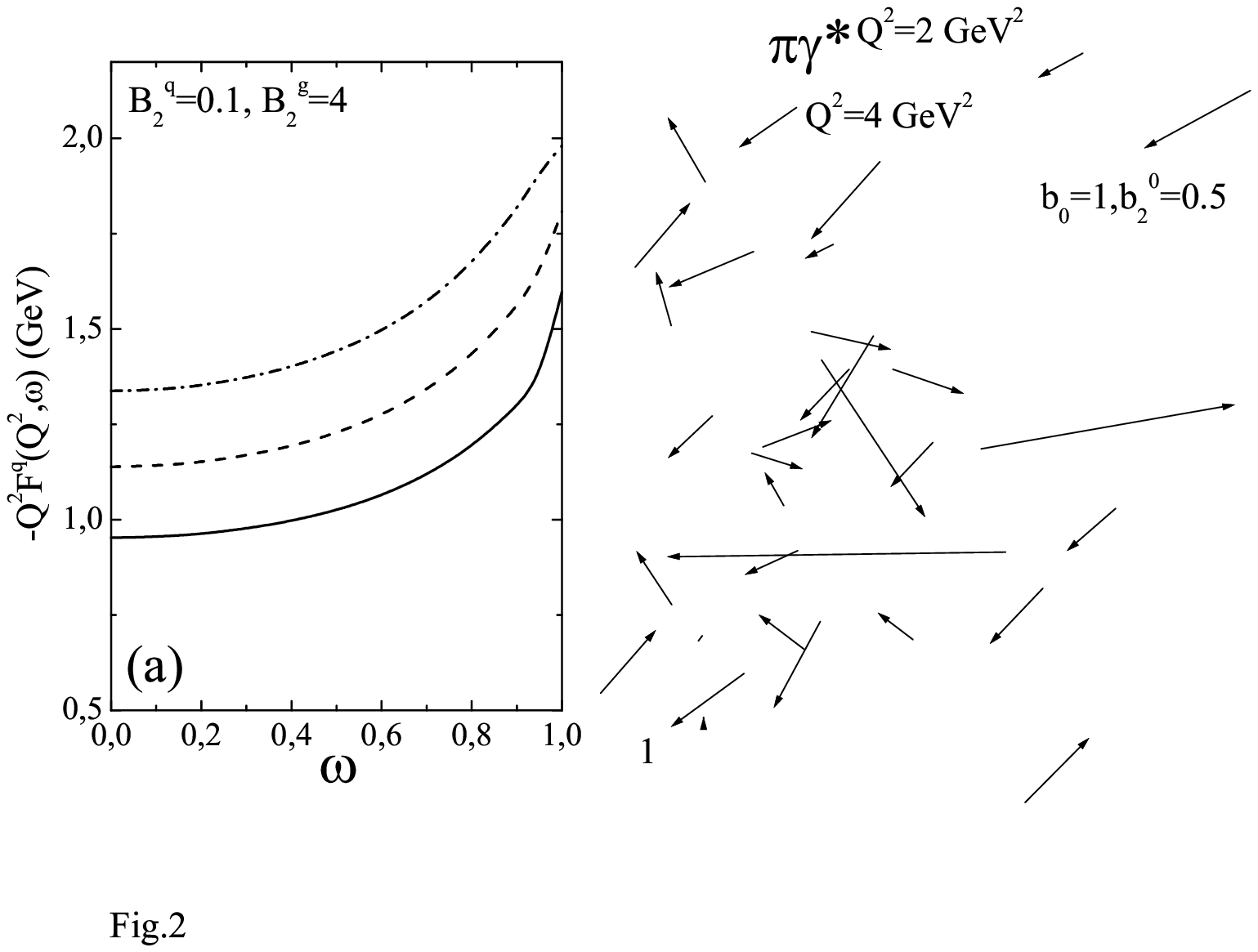,height=8cm,width=6.0cm,clip=}
\hspace{1.5cm}
\centering\epsfig{file=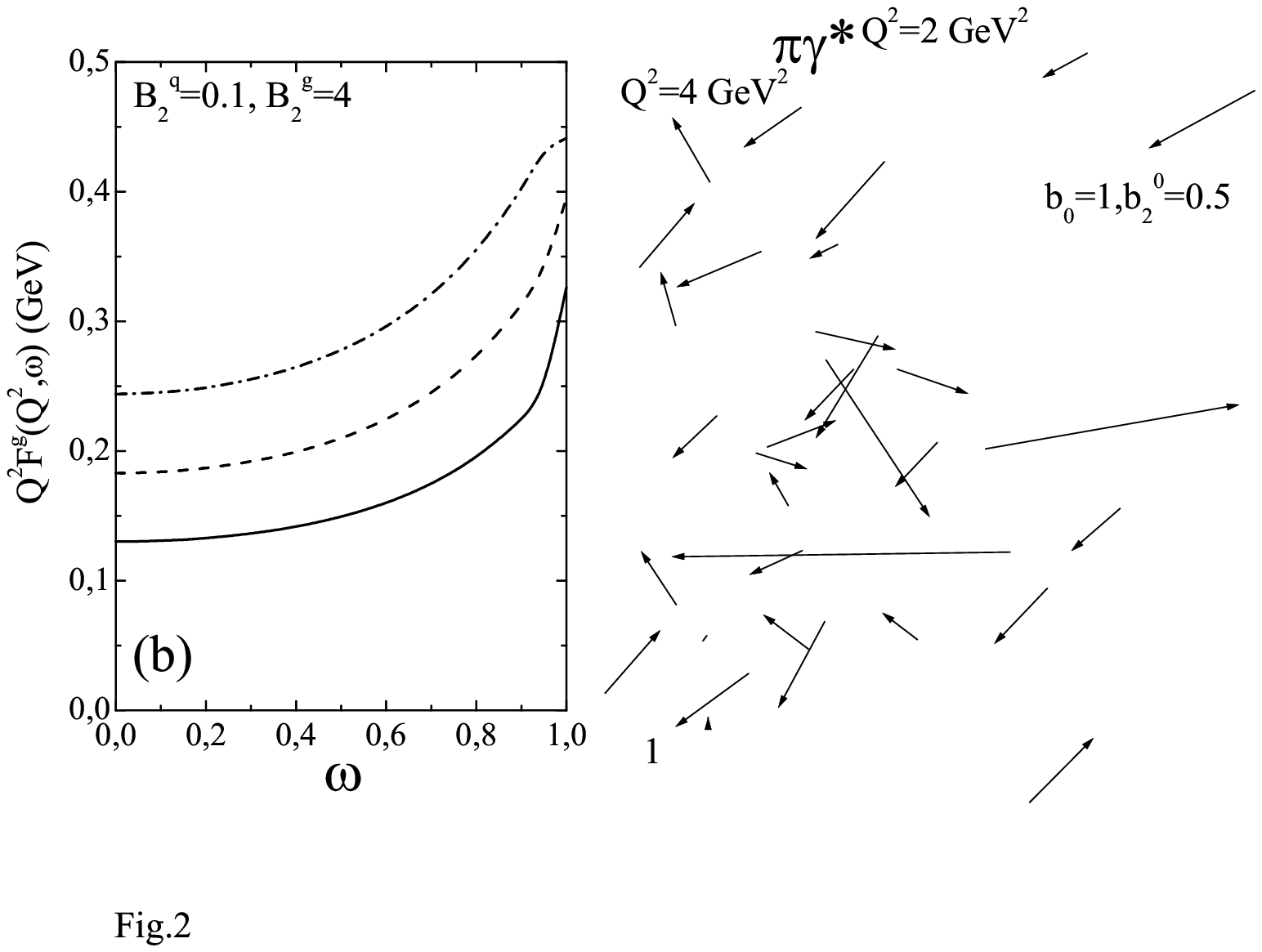,height=8cm,width=6.0cm,clip=}
\vspace{0.0cm}
\caption{The quark a) and gluon b) components of the form factor
$Q^2F_{\eta^{\prime}g^{*}g^{*}}(Q^2,\omega)$
at fixed
$Q^2$ vs. $\omega$. The RC method is used. The momentum scale for the
solid curves $Q^2=10\; \rm {GeV}^2$, for the dashed ones
$Q^2=6\; \rm {GeV}^2$, and for the dot-dashed curves
$Q^2=4\; \rm {GeV}^2$.
}
\label{fig:gfig8}
\end{figure}
and gluon components of the space-like vertex function, the total
vertex function $Q^2F_{\eta^{\prime}gg^{*}}(Q^2,\omega=\pm 1)$ for
$B_2^g \neq 0$ lies below the asymptotic one (Fig.\ 6a).
Comparing the predictions derived within the RC method with those
following from the standard HSA (Fig.\ 6b), we see a quantitative
difference between corresponding curves.
\begin{figure}[t]
\centering\epsfig{file=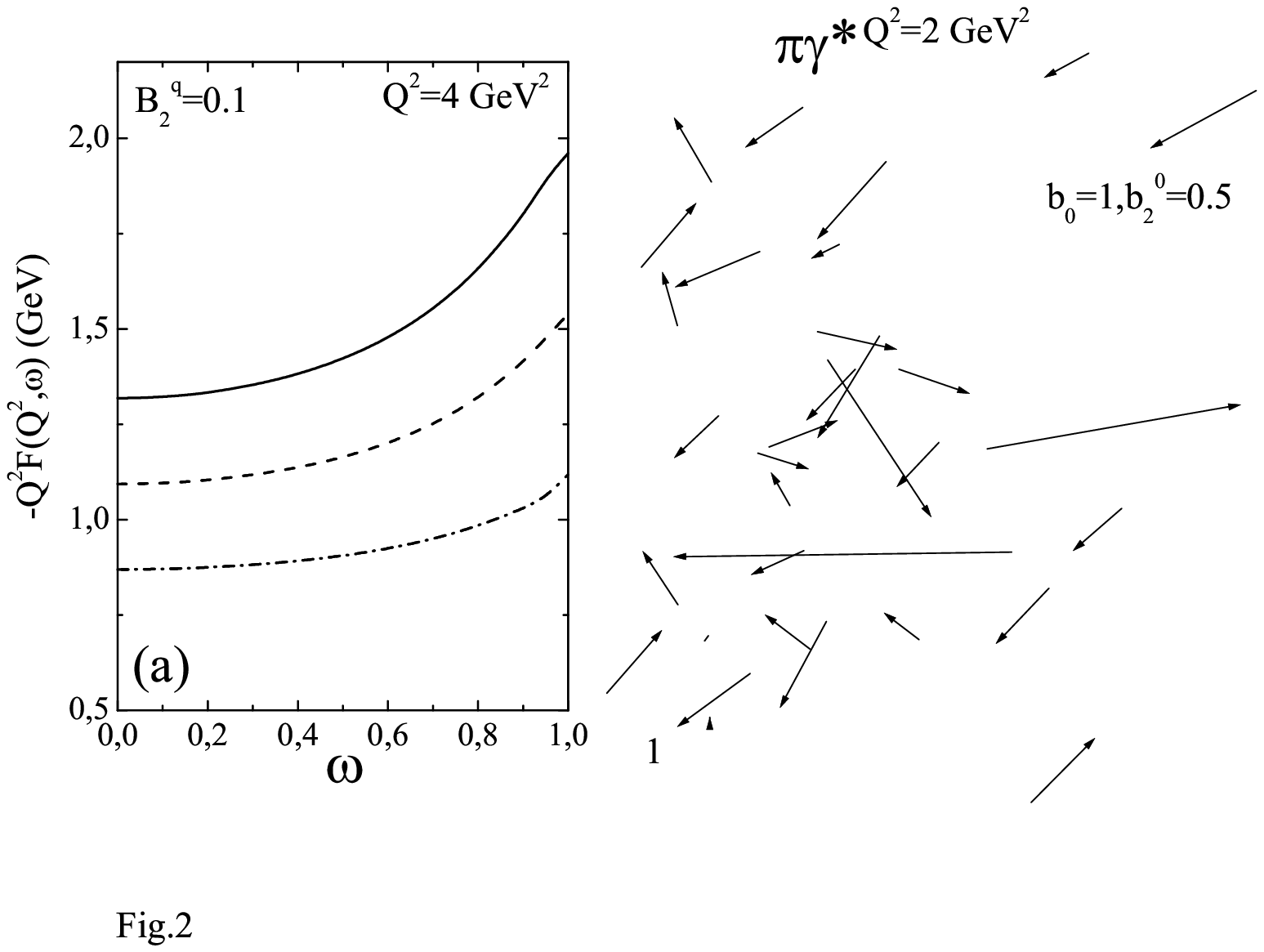,height=8cm,width=6.0cm,clip=}
\hspace{1.5cm}
\centering\epsfig{file=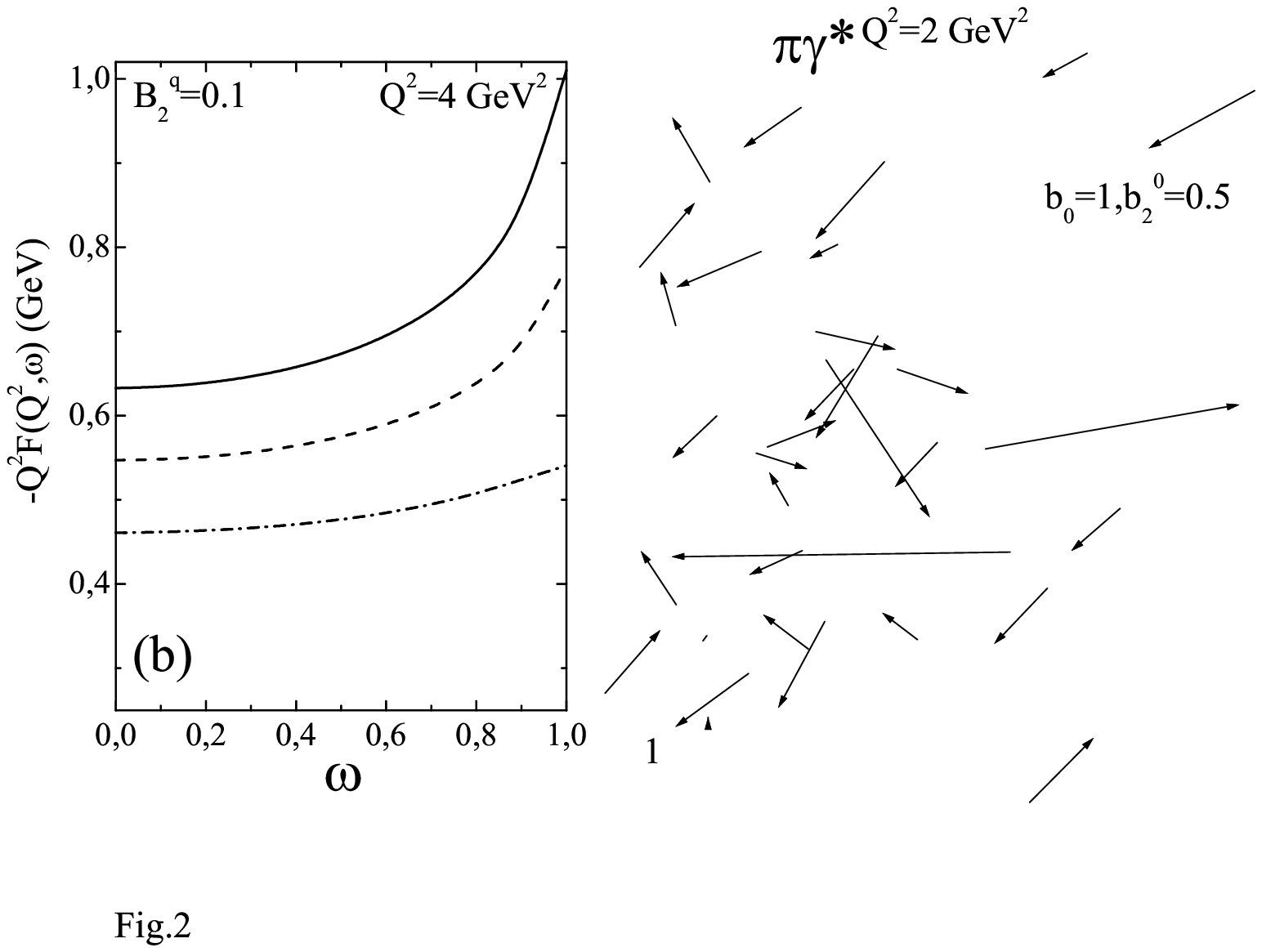,height=8cm,width=6.0cm,clip=}
\vspace{0.0cm}
\caption{ The FF $-Q^2F_{\eta^{\prime}g^{*}g^{*}}(Q^2,\omega)$
obtained by employing the RC method a) and the standard HSA b) as a
function of $\omega$. The expansion coefficients are $B_2^g=0$ for the
solid lines, $B_2^g=4$ for the dashed lines, and $B_2^g=8$ for the
dot-dashed lines.
}
\label{fig:gfig9}
\end{figure}

The dependence of the quark and gluon components of the form factor
on the asymmetry parameter $\omega$ at fixed $Q^2$ and for
various DA's of the $\eta^{\prime}$-meson are shown in
Fig.\ \ref{fig:gfig7}.
\begin{figure}[t]
\centering\epsfig{file=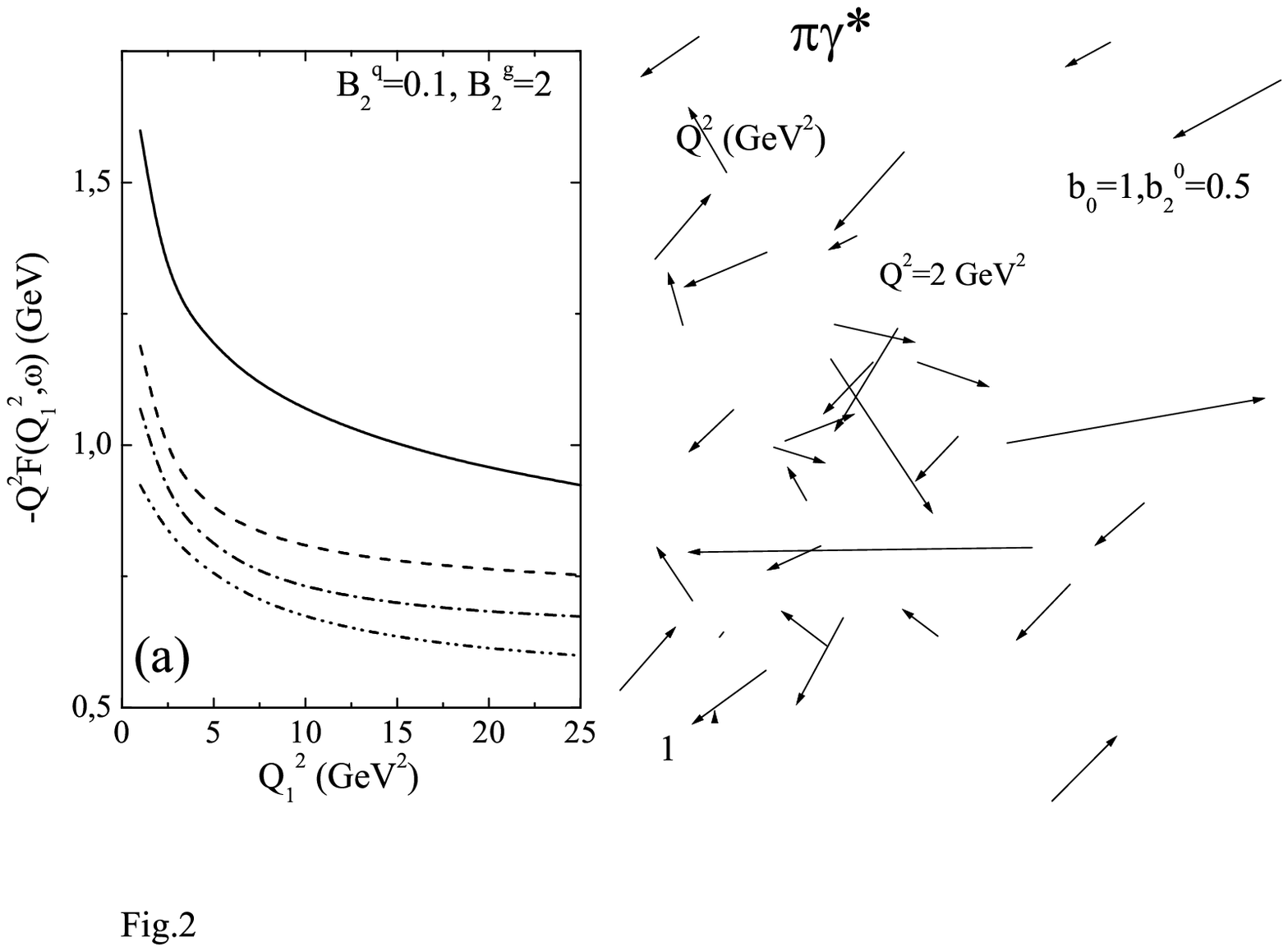,height=8cm,width=6.0cm,clip=}
\centering\epsfig{file=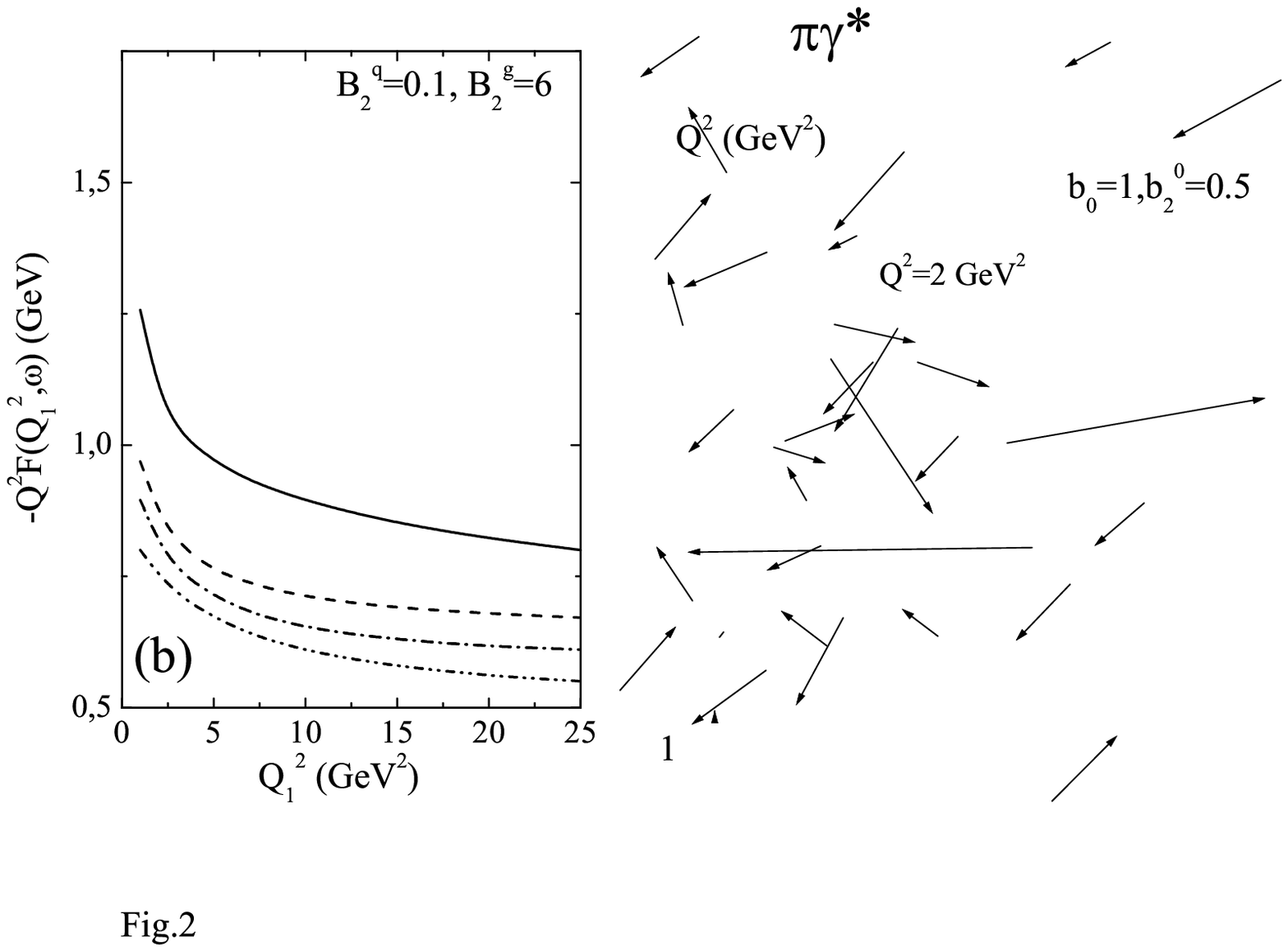,height=8cm,width=6.0cm,clip=}
\vspace{0.0cm}
\caption{ The form factor
$-Q^2F_{\eta^{\prime}g^{*}g^{*}}(Q_1^2,\omega)$ for two different DAs
of the $\eta^{\prime}$ meson and at different fixed values of $Q_2^2$;
viz., for the solid curves $Q_2^2=1\; \rm {GeV}^2$, for the dashed
curves
$Q_2^2=5\; \rm {GeV}^2$, for the dot-dashed curves
$Q_2^2=10\; \rm {GeV}^2$, and for the dot-dot-dashed curves
$Q_2^2=25\; \rm {GeV}^2$.
}
\label{fig:gfig10}
\end{figure}
Because $F_{\eta^{\prime}g^{*}g^{*}}(Q^2,\omega)$ is symmetric under
the exchange $\omega \leftrightarrow -\omega$, we present our
results in the region $0 \leq \omega \leq 1$ only.
We have just demonstrated in Fig.\ \ref{fig:gfig5} that the quark
component of the FF $F^q_{\eta^{\prime}gg^{*}}(Q^2,\omega=\pm 1)$ is
not sensitive to the use of various $\eta^{\prime}$-meson DA's,
employed in our calculations.
This is valid also for its behavior as a function of $\omega$
(Fig.\ 7a).
In accordance with our computations, an effect of the chosen parameter
$B_2^g$ on the gluon component of the FF is considerable in the
whole range of $\omega \in [0,1]$.

The magnitude of the quark and gluon components of the form factor
for a given DA depends on the total gluon virtuality $Q^2$
(Fig.\ \ref{fig:gfig8}).
In this case both the quark and gluon contributions to the FF
demonstrate sensitivity to the fixed value of $Q^2$.

The features of the quark and gluon components of the
$\eta^{\prime}g^{*}$ transition FF described above determine the
behavior of their sum as a function of the asymmetry parameter
$\omega$ and, as a result, we get the picture shown in Fig.\ 9a.
Owing to the gluon component, the form factor
$F_{\eta^{\prime}g^{*}g^{*}}(Q^2,\omega)$
depends on the $\eta^{\prime}$-meson DA used in the calculations.
For comparison, in Fig.\ 9b the curves found within the standard HSA
are also shown.
An enhancement of about a factor of two of the vertex function due to
power corrections is evident.

The $\eta^{\prime}g^{*}$ transition FF as function of the first gluon
virtuality $Q^2_1$ at various fixed values of the second one, $Q_2^2$,
and for different DA's is plotted in Fig.\ \ref{fig:gfig10}.

\begin{figure}[t]
\centering\epsfig{file=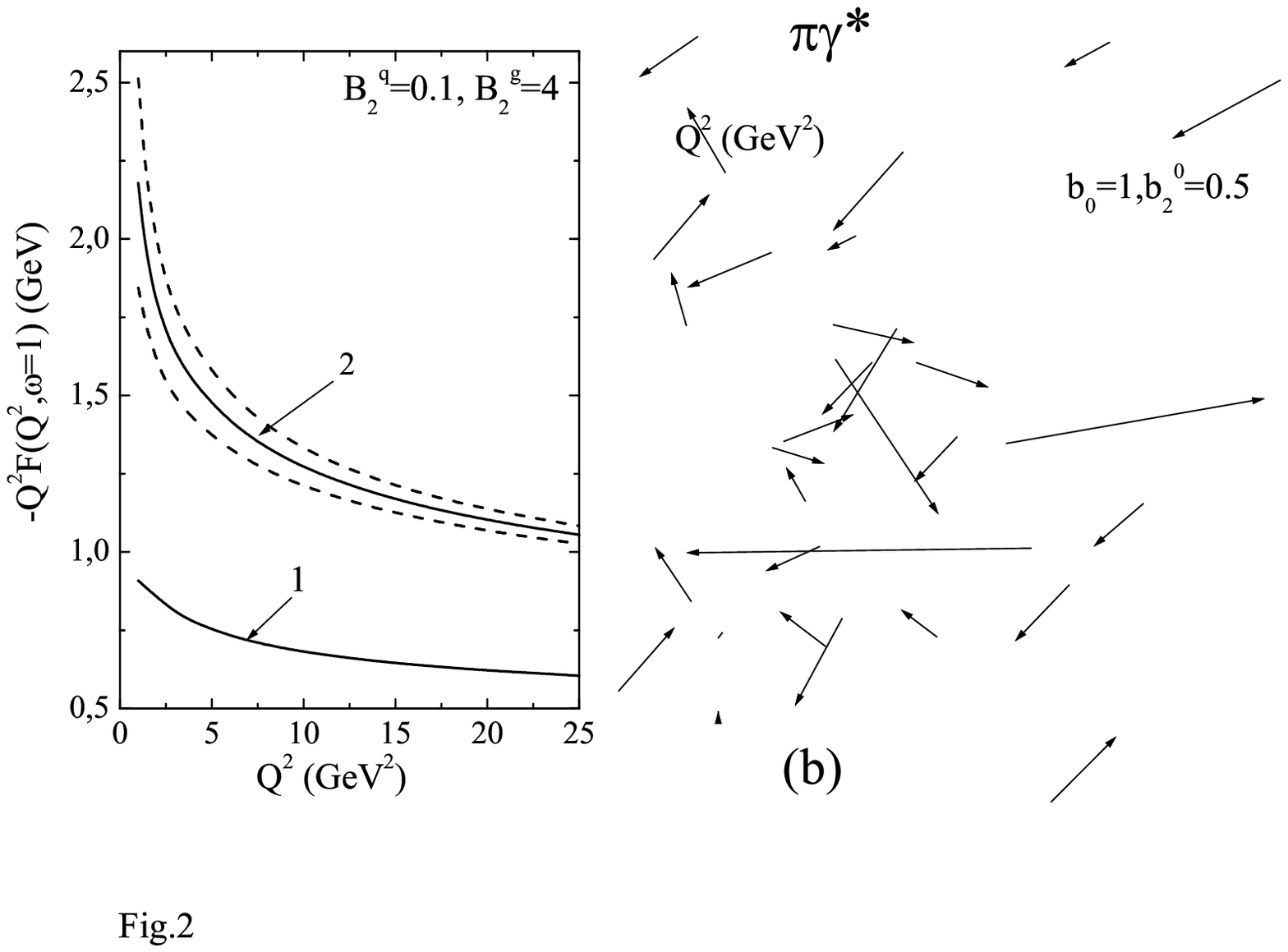,height=8cm,width=6.0cm,clip=}
\hspace{1.5cm}
\caption{
Influence of higher-twist ambiguities on the form factor
$-Q^{2}F_{\eta^{\prime}gg^{*}}(Q^{2},\omega=\pm 1)$,
calculated with the RC method (broken lines).
The solid lines, labelled $1$ and $2$, correspond to the FF found before
within perturbative QCD and the RC method, respectively.
For the upper dashed line the constants determining the ambiguities are
taken to be equal to $\{ N_{q} \}=1$, while for the lower dashed line
$\{ N_{q} \}=-1$, where $q=1, \: 2$ and $3$.
}
\label{fig:gfig11}
\end{figure}

As we have noted in Sect.\ III, the principal value prescription,
adopted in this work to regularize divergent integrals, produces
higher-twist ambiguities.
It is important to clarify to what extent these ambiguities may
alter our predictions.
To illustrate this effect, we depict in Fig.\ \ref{fig:gfig11}, as an
example, the scaled $\eta^{\prime}g$ transition FF
$Q^2F_{\eta^{\prime}gg^{*}}(Q^2,\omega=\pm 1)$.
The higher-twist ambiguities lead approximately to $\pm 15 \%$ shifts
relative to the previous RC result (line $2$).
On the other hand, comparing the standard pQCD prediction (solid line
1) with the lower dashed line (that denotes the RC prediction with a
negative higher-twist ambiguity), we see that still the power
corrections provide an enhancement of the standard pQCD result by a
factor $\sim 2$ in the region $Q^2 \sim 1-2\ {\rm GeV}^2$ and by a
factor $\sim 1.6$ at $Q^2=25\ {\rm GeV}^2$.
This latter effect is connected not only to larger uncertainties, but
also reflects the general trend of the FF to reach its asymptotic
value, i. e., the line $1$ in the limit $Q^2 \to \infty$.

\section{Concluding remarks}

In this paper we have evaluated power-suppressed corrections $\sim
1/Q^{2n},\;n=1,2\ldots$ to the space-like $\eta^{\prime}$-meson
virtual-gluon transition form factor
$Q^2F_{\eta^{\prime}g^{*}g^{*}}(Q^2,\omega)$.
To this end, we have employed the standard hard-scattering approach and
the running-coupling method in conjunction with the infrared-renormalon
calculus.
In the calculations, both the quark and the gluon distribution
amplitudes of the $\eta^{\prime}$ meson have been taken into account.
In these model DA's only the first non-asymptotic terms have been
retained and the values of the input coefficients
$B_2^q(1\ {\rm GeV}^2)$ and $B_2^g(1\ {\rm GeV}^2)$ extracted from the
analysis of the CLEO data on the $\eta^{\prime}\gamma$ electromagnetic
transition FF have been employed.

In order to apply the RC method to the considered process, the
hard-scattering amplitudes of the corresponding subprocesses
have been generalized in such a way as to preserve the symmetry
properties of both the hard-scattering amplitudes and the transition
form factor itself under the replacements
$x\leftrightarrow \overline{x}$ and $\omega \leftrightarrow
-\omega$.
In the computations within the RC method, the Laplace transformed
expression for the running coupling has been employed.
The Borel resummed form factors, obtained this way, have been
regularized by means of the principal value prescription.
Various limits of the general expression
$Q^2F_{\eta^{\prime}g^{*}g^{*}}(Q^2,\omega)$ have been found.

Our expressions for the vertex function
$F_{\eta^{\prime}g^{*}g^{*}}(Q^2,\omega)$, found within the
standard HSA, are in agreement with corresponding predictions made
in Ref.\ \cite{pas} (up to a conventional sign factor).
It has been demonstrated that (see Eq.\ (\ref{eq:2.31})) both the
quark and the gluon components of the vertex function are symmetric
under the exchange $\omega \leftrightarrow -\omega$, as they must
be owing to the Bose symmetry of the final gluons.
The RC method has provided us a tool for estimating power corrections
to the vertex function
$F_{\eta^{\prime}g^{*}g^{*}}(Q^2,\omega)$.
These corrections are implicitly contained in the pQCD factorization
formulas (\ref{eq:2.16}), (\ref{eq:2.17}) and originate from the
end-point $x \to 0;1$ regions.
It is clear that such corrections cannot been taken into account
in the standard pQCD approach by freezing the renormalization scale
$\mu_{R}^2$ and ignoring its dependence on the longitudinal momentum
fraction $x$.
As an important consistency check, we have proven that the results
obtained with the RC method in the asymptotic limit $Q^2 \to \infty$
reproduce the standard pQCD predictions for the vertex function.
This provides further justification for the treatment of the
hard-scattering amplitudes (\ref{eq:2.33}), (\ref{eq:2.34}), and the
symmetrization procedure employed in the running coupling method.

The presented numerical analysis shows that power corrections
considerably enhance the standard pQCD predictions for the form factor
in the explored region $1\;\rm {GeV}^2 \leq Q^2 \leq 25\; \rm {GeV}^2$,
though other sources, not considered here, may also give rise to
power corrections.
Our investigations demonstrate that the quark component of the form
factor at fixed $B_2^q=0.1$ is practically stable for several values
of $B_2^g=0,\;2,\;4,\;6,\; 8$.
Contrary to this, the gluon component of the FF is sensitive to the
adopted value of $B_2^g$.
As a consequence, the $\eta^{\prime}g^{*}$ transition FF was found to
depend on the gluonic content of the $\eta^{\prime}$ meson.
In the considered region the gluon contribution reduces the absolute
value of the space-like form factor
$Q^2F_{\eta^{\prime}g^{*}g^{*}}(Q^2,\omega)$.

As is true of any technique for calculating power corrections,
the theoretical framework elaborated in the present work makes
assumptions about the regularization of endpoint divergences.
There are of course other possibilities.
Nevertheless, we believe that our method is useful in pQCD
analyses of the $B$ meson exclusive decays and heavy-to-light
$B \to \pi, \rho$ transition form factors in the domain of moderate
momentum transfers.
A generalization of the RC method to describe time-like transitions
as well as its combination with resummation techniques to include
Sudakov logarithms will be the subject of separate investigations.
\bigskip
\\

{\Large {\bf Acknowledgments}}
\bigskip

One of the authors (S.S.A.) would like to thank Prof.~S.~Randjbar-Daemi
and the High Energy Section members for their hospitality at the Abdus
Salam ICTP, where this work was started and Prof.\ K.\ Goeke and the members
of the Institute for Theoretical Physics II, where this investigation was
completed.
Both of us wish to thank M.\ Polyakov, A.\ Bakulev, A.\ Belitsky, and
K.\ Passek-Kumeri\v{c}ki for useful comments.
The financial support by DAAD (S.S.A.) is gratefully acknowledged.

\newpage
\begin{appendix}
\appendix
\section{}
\renewcommand{\theequation}{\thesection.\arabic{equation}}
  \label{sec:appendix}\setcounter{equation}{0}

  In this Appendix we collect expressions for the $\eta^{\prime}$-meson
quark and gluon DA's for $n_f=4$. The parameters  (\ref{eq:2.42})
which determine these DA's assume the following values
$$
\gamma_{gg}^2=-\frac{35}{3},\; \gamma_{qg}^2=4, \; \gamma_{+}^2 \simeq -\frac{48}{9},
\; \gamma_{-}^2 \simeq -\frac{107}{9},
$$
$$
\rho_2^q \simeq \frac{19}{5}, \; \rho_2^g \simeq -\frac{1}{102}.
$$

It is evident that the elements of the anomalous dimensitions matrix
$\gamma_{qq}^2$ and $\gamma_{gq}^2$ are $n_f$ independent and have the
values shown in Eq.\ (\ref{eq:2.42}).

In the $n_f=4$ case the DA's $\phi^q(x,Q^2)$ and $\phi^g(x,Q^2)$ of the
$\eta^{\prime}$-meson have the form (\ref{eq:2.43}) as well, the only difference
being in the functions $A(Q^2)$ and $B(Q^2)$, which now are defined
by the expressions
$$
A(Q^2)=6B_2^q \left ( \frac{\alpha_{\rm s}(Q^2)}{\alpha_{\rm s}(\mu_0^2)}
\right )^{\frac{48}{75}}-\frac{B_2^g}{17} \left (\frac{\alpha_{\rm
s}(Q^2)}{\alpha_{\rm s}(\mu_0^2)}\right )^{\frac{107}{75}}
$$

\begin{equation}
\label{eq:1app}
B(Q^2)=19B_2^q \left ( \frac{\alpha_{\rm s}(Q^2)}{\alpha_{\rm s}(\mu_0^2)}
\right )^{\frac{48}{75}}+5B_2^g \left (\frac{\alpha_{\rm
s}(Q^2)}{\alpha_{\rm s}(\mu_0^2)}\right )^{\frac{107}{75}}.
\end{equation}

In the previous sections all results have been written down for $n_f=3$
valid for a momentum transfers $ 1\ {\rm GeV}^2 \leq \ Q^2 \ <2\ {\rm GeV}^2 $.
For momentum transfers in the range
$ 2\ {\rm GeV}^2\ \leq  Q^2 \ \leq 25\ {\rm GeV}^2 $,
the choice $n_f=4$ has to be employed. In this case, the expressions that
determine the quark component of the
transition FF remain unchanged, except for the function $A(Q^2)$ (and
$\beta_0$, $R(u,t)$), which
should be taken from Eq.\ (\ref{eq:1app}). The expressions for the
gluon component of the FF have to be rescaled by a factor $3/4$---
apart from the replacement of the function $B(Q^2)$ (and
$\beta_0$, $R(u,t)$)---since the hard-scattering
amplitudes $T_{1(2)}^g(x,Q^2,\omega)$ explicitly depend on $n_f$
(\ref{eq:2.34}).
\end{appendix}

\end{document}